\documentclass[prd,floatfix,showpacs]{revtex4}
\usepackage{epsfig}

\newcommand{\eq}[1]{Eq.~(\ref{#1})}

\newcommand{\be}{\begin{equation}}
\newcommand{\ee}{\end{equation}}
\newcommand{\bea}{\begin{eqnarray}}
\newcommand{\eea}{\end{eqnarray}}
\newcommand{\ben}{\begin{eqnarray*}}
\newcommand{\een}{\end{eqnarray*}}

\newcommand{\DS}{Dyson--Schwinger }

\newcommand{\ST}{Slavnov--Taylor }
\newcommand{\YM}{Yang--Mills }

\newcommand{\w}{\omega}
\newcommand{\e}{\varepsilon}
\newcommand{\al}{\alpha}
\newcommand{\ba}{\beta}
\newcommand{\ga}{\gamma}
\newcommand{\G}{\Gamma}
\newcommand{\de}{\delta}

\newcommand{\si}{\sigma}
\newcommand{\Si}{\Sigma}
\newcommand{\ro}{\rho}
\newcommand{\la}{\lambda}

\newcommand{\ka}{\kappa}
\newcommand{\ta}{\tau}
\newcommand{\et}{\eta}
\newcommand{\ha}{\frac{1}{2}}
\newcommand{\pd}{\partial}

\renewcommand{\th}{\theta}
\newcommand{\cd}{{\cal D}}
\newcommand{\cs}{{\cal S}}
\renewcommand{\div}{\vec{\nabla}}

\newcommand{\s}[2]{{#1}\!\cdot\!{#2}}
\newcommand{\ov}[1]{\overline{#1}}
\newcommand{\dk}[1]{\,\,\,\raisebox{-0.4ex}{\large $\bar{}$}\!\!d\,{#1}\,}
\newcommand{\dx}[1]{d^4{#1}\,}

\newcommand{\ev}[1]{<\!\!{#1}\!\!>}
\begin{document}
\title{Slavnov-Taylor identities in Coulomb gauge Yang-Mills theory}
\author{P.~Watson}
\author{H.~Reinhardt}
\affiliation{Institut f\"ur Theoretische Physik, Universit\"at T\"ubingen, 
Auf der Morgenstelle 14, D-72076 T\"ubingen, Deutschland}
\begin{abstract}
The Slavnov-Taylor identities of Coulomb gauge Yang-Mills theory are 
derived from the (standard, second order) functional formalism.  It is shown 
how these identities form closed sets from which one can in principle fully 
determine the Green's functions involving the temporal component of the 
gauge field without approximation, given appropriate input.
\end{abstract}
\maketitle
\section{Introduction}
\setcounter{equation}{0}

The issue of color confinement in quantum chromodynamics [QCD], widely 
accepted as the theory of the strong interaction, is a longstanding problem 
both conceptually and quantitatively.  One necessary prerequisite for color 
confinement is the conservation of color charge: a `leaky' system clearly 
cannot be confining.  Within the framework of functional field-theoretic 
methods in nonabelian theories (i.e., QCD), expressed in terms of local 
Green's functions, the Slavnov--Taylor identities 
\cite{Taylor:1971ff,Slavnov:1972fg} are the consequence of this charge 
conservation.

The Slavnov--Taylor identities come in various guises and have been applied 
to many problems, in particular to the nonperturbative study of the \DS 
equations.  For example, in quantum electrodynamics [QED] where the 
Slavnov--Taylor identities reduce to the Ward identity, their study led to 
the Ball--Chiu vertex \cite{Ball:1980ay} with its subsequent improvement, 
the Curtis--Pennington vertex \cite{Curtis:1990zs}, which have been used to 
study dynamical mass generation (see for example 
Refs.~\cite{Pennington:1998cj,Alkofer:2000wg}).  In axial gauge Yang--Mills 
theory, the Slavnov--Taylor identities \cite{Kummer:1974ze} have been used 
to study the gluon propagator \DS equation, with the inference that the 
propagator diverges as $1/q^4$ in the infrared (one potential signal for 
confinement) \cite{Baker:1980gf}.  This result is however, not without 
ambiguity \cite{West:1982gg}.  In Landau gauge Yang--Mills theory, the 
Mandelstam approximation \cite{Mandelstam:1979xd} (see also 
Ref.~\cite{BarGadda:1979cz}), where ghost contributions to the three-gluon 
vertex Slavnov--Taylor identity are neglected, one can also obtain an 
infrared enhanced gluon propagator.  The above examples can be colloquially 
referred to as Schwinger--Dyson studies (these are early studies and a quirk 
of history has subsequently resulted in the reordering of the names) and 
they share two common features: the utilization of the Slavnov--Taylor 
identities and either require no, or assume no ghost contributions.

In attempting to further the Mandelstam approximation to Landau gauge 
Yang--Mills theory by including the ghost contributions, a startling new 
phenomenon emerged --- infrared ghost dominance \cite{von Smekal:1997vx} 
(see also \cite{Alkofer:2000wg,Fischer:2006ub} for contemporary reviews).  
The starting point was an approximation to the Slavnov--Taylor identity for 
the ghost-gluon vertex.  This approximation was shown to be inconsistent 
with perturbation theory \cite{Watson:1999ha}, but this turned out to be 
unimportant: using general arguments one can show that the tree-level part 
of the ghost-gluon vertex is the important ingredient 
\cite{Watson:2001yv}.  Indeed, there are indications that the dressing of 
the ghost-gluon vertex function is only slight 
\cite{Schleifenbaum:2004id}.  The results of this infrared ghost dominance 
can be summarized as follows: the gluon propagator is infrared suppressed, 
whereas the ghost propagator is infrared enhanced.  In the context of 
confinement, the results imply positivity violation (formalized in the 
Oehme--Zimmermann superconvergence relations \cite{Oehme:1979ai}) and are 
in agreement with the Kugo--Ojima \cite{Kugo:1979gm} and Gribov--Zwanziger 
\cite{Gribov:1977wm,Zwanziger:1995cv,Zwanziger:1998ez} confinement 
scenarios.  These results have been contested (see for example 
Ref.~\cite{Boucaud:2008ky}) by appealing to the Slavnov--Taylor identity 
for the three-gluon vertex, applying the further assumption that certain 
dressing functions that compose the ghost-gluon vertex kernel remain 
infrared finite and with the conclusion that the ghost propagator is not 
infrared enhanced.  The existence of such additional solutions has been 
verified in Ref.~\cite{Fischer:2008uz} (this type of solution was 
incidentally also observed in Ref.~\cite{Epple:2007ut} within the canonical 
approach to Coulomb gauge Yang--Mills theory \cite{Feuchter:2004mk}): it was 
concluded however that the Slavnov--Taylor identities alone cannot 
discriminate between the two types of solution and that one needs explicit 
additional input (the authors arguing that the infrared enhanced ghost 
solution is preferred).  

From the above (admittedly historically incomplete discussion of) works, 
several lessons emerge.  Firstly, the Slavnov--Taylor identities play a 
central role in nonperturbative \DS studies since they provide information 
about the higher $n$-point functions that enter the \DS equations.  Second, 
in all the above studies, the Slavnov--Taylor identities relate the 
$4$-vector contraction of a vertex to some combination of inverse 
propagators (proper two-point functions).  This contraction means that only 
part of the vertex is constrained by the identity and it is not necessarily 
that part which enters the \DS equations.  One either assumes that the 
`transverse' vertex (i.e., that part not constrained by the Slavnov--Taylor 
identity) can be neglected or considers more sophisticated input to complete 
the closure of the system such that the \DS equations can be solved (e.g., 
the Curtis--Pennington vertex uses multiplicative renormalizability 
constraints \cite{Curtis:1990zs}).  Third, even if the Slavnov--Taylor 
identities are applied to close the \DS equations, the solution (and in 
particular the infrared behavior) may not be uniquely specified and requires 
additional input, as demonstrated in Ref.~\cite{Fischer:2008uz}.  These 
latter two points are emphatically not intended as a criticism of the \DS 
approach, merely as reminders of some of the problems encountered 
(confinement is after all, not entirely trivial) and in this paper, we will 
see what Coulomb gauge has to say about the subject.

The importance of Coulomb gauge in studying nonperturbative QCD was 
recognized early on, as was the inherent difficulty in technical calculation 
for such noncovariant gauges \cite{Abers:1973qs}.  The significance of 
Coulomb gauge is based in the observation that in this gauge, the system 
reduces naturally to physical degrees of freedom (explicitly demonstrated 
in Ref.~\cite{Zwanziger:1998ez}).  Given that the Slavnov--Taylor identities 
are the expression of charge conservation as applied to Green's functions, 
there exists a clear motivation to derive them in Coulomb gauge.  In 
addition, there has recently been much technical progress in the Coulomb 
gauge functional formalism that provides the background to the study of the 
\DS equations: the derivation of the \DS equations themselves 
\cite{Watson:2006yq,Watson:2007vc}, their one-loop perturbative analysis 
\cite{Watson:2007mz,Watson:2007vc,Popovici:2008ty}, the resolution of more 
formal aspects of the (incomplete) gauge-fixing and the existence of a 
conserved, vanishing total charge (and the absence of the infamous Coulomb 
gauge energy divergences) \cite{Watson:2007fm,Reinhardt:2008pr}.  With the 
Slavnov--Taylor identities, most of the components required for at least an 
initial study of the \DS equations are in place.

In this paper, we thus derive the Slavnov--Taylor identities of Coulomb 
gauge Yang--Mils theory in the second order functional formalism and discuss 
some of their immediate consequences.  In Section~2, the basic formulation 
of the functional formalism and the Gauss--BRST invariance is presented.  
The Slavnov--Taylor identities for the two-point functions are derived in 
detail in Section~3.  In Section~4, the identities for the vertex functions 
are derived and it is shown how these identities form closed sets such that 
the temporal Green's functions may be obtained as their solution.  Section~5 
is concerned with the possibility of extracting unambiguous information 
about Green's functions from the identities in the infrared.  The paper 
concludes with a discussion of various aspects of the identities.  For the 
convenience of the reader, selected relevant results from earlier works are 
listed in Appendix~\ref{app:decomp}.  Lengthy configuration space 
expressions, important to the derivation of the Slavnov--Taylor identities 
but not to the narrative of the paper are relegated to 
Appendix~\ref{app:eqlist}.

\section{Functional formalism and Gauss--BRST invariance}
\setcounter{equation}{0}

Let us begin by considering Yang--Mills theory in the functional formalism.  
We will use the notation and conventions established in 
\cite{Watson:2006yq,Watson:2007vc}.  We work in Minkowski space with metric 
$g_{\mu\nu}=\mbox{diag}(1,-\vec{1})$.  Roman subscripts ($i,j,\ldots$) 
denote spatial indices and superscripts ($a,b,\ldots$) denote color 
indices.  We will often write configuration space coordinates ($x,y,\ldots$) 
as subscripts where no confusion arises.

The \YM action is defined as
\be
\cs_{YM}=\int\dx{x}\left[-\frac{1}{4}F_{\mu\nu}^aF^{a\mu\nu}\right]
\ee
where the (antisymmetric) field strength tensor $F$ is given in terms of the 
gauge field $A_{\mu}^a$:
\be
F_{\mu\nu}^a
=\pd_{\mu}A_{\nu}^a-\pd_{\nu}A_{\mu}^a+gf^{abc}A_{\mu}^bA_{\nu}^c.
\ee
In the above, the $f^{abc}$ are the structure constants of the $SU(N_c)$ 
group whose generators obey $\left[T^a,T^b\right]=\imath f^{abc}T^c$.  The 
Yang-Mills action is invariant under a local $SU(N_c)$ gauge transform 
characterized by the parameter $\th_x^a$:
\be
U_x=\exp{\left\{-\imath\th_x^aT^a\right\}}
\ee
such that for infinitesimal $\th_x^a$, the gauge field transforms as
\be
A_{\mu}^a\rightarrow A_{\mu}^{\prime a}
=A_{\mu}^a-\frac{1}{g}\hat{D}_{\mu}^{ab}\th^b
\ee
with the covariant derivative in the adjoint color representation given by
\be
\hat{D}_{\mu}^{ac}=\de^{ac}\pd_{\mu}+gf^{abc}A_{\mu}^b.
\ee
In terms of the temporal and spatial components, the above transform reads 
(we rewrite the temporal component $A_0$ as $\si$)
\bea
\si^a\rightarrow\si^{\prime a}
&=&\si^a-\frac{1}{g}\pd_{0}\th^a-f^{abc}\si^b\th^c,\nonumber\\
\vec{A}^a\rightarrow\vec{A}^{\prime a}
&=&\vec{A}^a+\frac{1}{g}\div\th^a-f^{abc}\vec{A}^b\th^c.
\eea

Consider the functional integral
\be
Z=\int\cd\Phi\exp{\left\{\imath\cs_{YM}\right\}}
\ee
where $\cd\Phi$ denotes the functional integration measure for the 
collection of all fields.  Since the action is invariant under gauge 
transformations, $Z$ is divergent by virtue of the integration over the 
gauge group.  To overcome this problem we use the Faddeev-Popov technique 
and introduce a gauge-fixing term along with an associated ghost term 
\cite{IZ}.  Using a Lagrange multiplier field, $\la^a$, to implement the 
gauge-fixing, in Coulomb gauge ($\s{\div}{\vec{A}}^a=0$) we can then write
\be
Z=\int\cd\Phi\exp{\left\{\imath\cs_{YM}+\imath\cs_{FP}\right\}},\;\;\;\;
\cs_{FP}=\int d^4x\left[-\la^a\s{\vec{\nabla}}{\vec{A}^a}
-\ov{c}^a\s{\vec{\nabla}}{\vec{D}^{ab}}c^b\right]
\label{eq:zgf}
\ee
where $\ov{c}^a$ and $c^b$ are the Grassmann-valued ghost fields.  The new 
term in the action, $\cs_{FP}$, is invariant under the 
Gauss--Becchi--Rouet--Stora--Tyutin [Gauss-BRST] transform 
\cite{Zwanziger:1998ez} whereby the infinitesimal, spacetime dependent 
gauge parameter $\th_x^a$ is factorized into two Grassmann-valued 
components: $\th_x^a=c_x^a\de\la_t$, where $\de\la_t$ is the 
\emph{time-dependent} infinitesimal variation (not to be confused with the 
colored Lagrange multiplier field $\la^a$).  The Gauss-BRST transform is 
peculiar to Coulomb gauge --- the time-dependent variation is allowed 
simply because the gauge-fixing does not involve any explicit 
time-derivatives.  The variations of the new fields read:
\be
\de\ov{c}_x^a=\frac{1}{g}\la_x^a\de\la_t,\;\;\;\;
\de c_x^a=-\ha f^{abc}c_x^bc_x^c\de\la_t,\;\;\;\;
\de\la_x^a=0.
\ee
By including a source term, the generating functional is given by
\be
Z[J]=
\int\cd\Phi\exp{\left\{\imath\cs_{YM}+\imath\cs_{FP}+\imath\cs_s\right\}}
\ee
where
\be
\cs_s=\int d^4x\left[\ro^a\si^a+\s{\vec{J}^a}{\vec{A}^a}+\ov{c}^a\et^a
+\ov{\et}^ac^a+\xi^a\la^a\right].
\ee
Regarding the Gauss-BRST transform as a change of integration variables 
under which the generating functional is invariant, noting that the 
associated Jacobian factor is trivial \cite{Watson:2006yq} and only the 
source term varies, we deduce that
\bea
0&=&\left.\int\cd\Phi\frac{\de}{\de\left[\imath\de\la_t\right]}
\exp{\left\{\imath\cs_{YM}+\imath\cs_{FP}
+\imath\cs_s+\imath\de\cs_s\right\}}\right|_{\de\la_t=0}
\nonumber\\
&=&\int\cd\Phi\exp{\left\{\imath\cs_{YM}+\imath\cs_{FP}+\imath\cs_s\right\}}
\int d^4x\de(t-x_0)\times
\nonumber\\&&
\left[-\frac{1}{g}(\pd_x^0\ro_x^a)c_x^a+f^{abc}\ro_x^a\si_x^bc_x^c
-\frac{1}{g}J_{ix}^a\nabla_{ix}c_x^a+f^{abc}J_{ix}^aA_{ix}^bc_x^c
+\frac{1}{g}\la_x^a\et_x^a
+\ha f^{abc}\ov{\et}_x^ac_x^bc_x^c\right].
\label{eq:wtid0}
\eea
This equation is the starting point for deriving the \ST identities.  Notice 
the $\de(t-x_0)$ constraint, which arises because of the time-dependent 
variation $\de\la_t$ and is characteristic to the Gauss-BRST transform.  
It leads eventually to a nontrivial energy injection into the \ST identities 
which is not present in the covariant gauge case.

It is pertinent at this stage to discuss some nontrivial points associated 
with the above.  Coulomb gauge is in fact not a complete gauge.  Even after 
adding the gauge-fixing terms, the functional integral, \eq{eq:zgf}, still 
contains zero-modes generated by purely temporal gauge transforms (and for 
that matter, global transforms too).  Explicitly separating these temporal 
zero-modes within the Faddeev--Popov procedure, one can formally show 
(within the first order formalism, but since this is connected to the 
second order formalism used here via identities, the same conclusions 
apply) that the total color charge of the system is constrained to be 
conserved and vanishing \cite{Reinhardt:2008pr}.  The above expression, 
\eq{eq:wtid0}, is the dynamical statement of color charge conservation --- 
it shows how the external sources must be arranged in accordance with the 
gauge symmetry of the underlying theory.  In the two differing contexts, 
the incompleteness of the gauge plays a central role: the formal isolation 
of the zero-modes gives the total charge constraint and the time-dependent 
variation ($\de\la_t$) will give (later) the extra temporal scale.  We will 
discuss this connection at the end.

Note also that the presence of the temporal zero-modes precludes deriving a 
similar expression to \eq{eq:wtid0} above by considering a purely 
time dependent (spatially independent) full gauge transform, i.e., 
$\th(t,\vec{x})\rightarrow\th(t)$, since the functional integration is 
ill-defined.  In such a case, the identity gives a resultant $\G_{\si\si}$ 
Green's function (see later for details of the definition) with a dressing 
function that vanishes at zero momentum and this would hold even at 
tree-level which clearly contradicts the perturbative behavior for which 
the dressing function is unity (see Ref.~\cite{Watson:2007vc} or 
Appendix~\ref{app:decomp}).  The reason for this is that when considering 
the gauge-variant Green's functions, the integration over the gauge group 
associated with the zero-modes averages such quantities to zero.  For the 
Gauss-BRST transform considered here, the spatial dependence of the 
ghost-field ensures that this problem is not encountered (since the 
gauge-fixing part of the action can be rewritten in terms of spatial 
derivatives of the ghost-field, they can always be implicitly defined as 
having no spatially constant component).

So far, the generating functional, $Z[J]$, generates all Green's functions, 
connected and disconnected.  The generating functional of connected Green's 
functions is $W[J]$ where
\be
Z[J]=e^{W[J]}.
\ee
We define the classical fields to be
\be
\Phi_\al=\frac{1}{Z}\int\cd\Phi\,\Phi_\al\exp{\imath\cs}
=\frac{1}{Z}\frac{\de Z}{\de\imath J_\al}
\ee
(we use the same notation for both the classical and quantum fields since no 
confusion will arise).  The generating functional of proper Green's 
functions is the effective action, $\G$, which is a functional of the 
classical fields and is defined through a Legendre transform of $W$:
\be
\G[\Phi]=W[J]-\imath J_\al\Phi_\al.
\label{eq:legtran}
\ee
In the above, we use a compact notation for the sources and fields: a 
generic field is denoted $\Phi_\al$, with source $J_\al$ and with the 
index $\al$ standing for all attributes of the field in question (including 
its type); further, summation over all discrete indices and integration over 
all continuous arguments is implicitly understood.  We introduce a bracket 
notation for derivatives of $W$ with respect to sources and of $\G$ with 
respect to classical fields (no confusion arises since the two sets of 
derivatives are never mixed):
\be
\ev{\imath J_\al}=\frac{\de W}{\de\imath J_\al},\;\;\;\;
\ev{\imath\Phi_\al}=\frac{\de\G}{\de\imath\Phi_\al}.
\ee
We can now rewrite \eq{eq:wtid0} as
\bea
0&=&\int\dx{x}\de(t-x_0)\left\{
\frac{1}{g}\left(\pd_x^0\ev{\imath\si_x^a}\right)c_x^a-f^{abc}
\ev{\imath\si_x^a}\si_x^bc_x^c+\frac{1}{g}
\ev{\imath A_{ix}^a}\nabla_{ix}c_x^a-f^{abc}
\ev{\imath A_{ix}^a}A_{ix}^bc_x^c-\frac{1}{g}\la_x^a\ev{\imath\ov{c}_x^a}
\right.\nonumber\\&&\left.
+\frac{1}{2}f^{abc}\ev{\imath c_x^a}c_x^bc_x^c-f^{abc}\ev{\imath\si_x^a}
\ev{\imath\ro_x^b\imath\ov{\et}_x^c}-f^{abc}\ev{\imath A_{ix}^a}
\ev{\imath J_{ix}^b\imath\ov{\et}_x^c}+\frac{1}{2}f^{abc}\ev{\imath c_x^a}
\ev{\imath\ov{\et}_x^b\imath\ov{\et}_x^c}
\right\}.
\label{eq:wtid1}
\eea
Knowing the functional form of the ghost equation of motion 
\cite{Watson:2007vc},
\be
\ev{\imath\ov{c}_x^a}=-\nabla_x^2c_x^a
+gf^{abc}\nabla_{ix}
\left[\ev{\imath J_{ix}^b\imath\ov{\et}_x^c}+A_{ix}^bc_x^c\right],
\label{eq:gheom0}
\ee
allows us to write
\bea
\lefteqn{
\int\dx{x}\de(t-x_0)\frac{1}{g}\ev{\imath A_{ix}^a}\nabla_{ix}c_x^a=
\int\dx{x}\de(t-x_0)\frac{1}{g}\left[\frac{\nabla_{ix}}{(-\nabla_x^2)}
\ev{\imath A_{ix}^a}\right]\nabla_{x}^2c_x^a}
&&\nonumber\\
&=&\int\dx{x}\de(t-x_0)\left\{-\frac{1}{g}
\left[\frac{\nabla_{ix}}{(-\nabla_x^2)}\ev{\imath A_{ix}^a}\right]
\ev{\imath\ov{c}_x^a}-f^{abc}\ev{\imath A_{ix}^a}
\frac{\nabla_{ix}\nabla_{jx}}{(-\nabla_x^2)}
\left[\ev{\imath J_{jx}^b\imath\ov{\et}_x^c}+A_{jx}^bc_x^c\right]\right\}.
\eea
Inserting the above into \eq{eq:wtid1} then gives
\bea
0&=&\int\dx{x}\de(t-x_0)\left\{
\frac{1}{g}\left(\pd_x^0\ev{\imath\si_x^a}\right)c_x^a-f^{abc}
\ev{\imath\si_x^a}\si_x^bc_x^c
-\frac{1}{g}\left[\frac{\nabla_{ix}}{(-\nabla_x^2)}
\ev{\imath A_{ix}^a}\right]\ev{\imath\ov{c}_x^a}
\right.\nonumber\\&&\left.
-f^{abc}\ev{\imath A_{ix}^a}t_{ij}(\vec{x})A_{jx}^bc_x^c
-\frac{1}{g}\la_x^a\ev{\imath\ov{c}_x^a}+\frac{1}{2}f^{abc}
\ev{\imath c_x^a}c_x^bc_x^c-f^{abc}\ev{\imath\si_x^a}
\ev{\imath\ro_x^b\imath\ov{\et}_x^c}
\right.\nonumber\\&&\left.
-f^{abc}\ev{\imath A_{ix}^a}t_{ij}(\vec{x})
\ev{\imath J_{jx}^b\imath\ov{\et}_x^c}+\frac{1}{2}f^{abc}
\ev{\imath c_x^a}\ev{\imath\ov{\et}_x^b\imath\ov{\et}_x^c}
\right\}
\label{eq:wtid2}
\eea
where $t_{ij}(\vec{x})=\de_{ij}+\nabla_{ix}\nabla_{jx}/(-\nabla_x^2)$ is 
the transverse projector in configuration space.

Before considering the full content of the \ST identities arising from 
\eq{eq:wtid2}, let us show that functional derivatives with respect to the 
Lagrange multiplier field, $\la_x^a$, play no further role.  The field 
equation of motion for $\la_x^a$, valid in the presence of sources reads 
\cite{Watson:2007vc}
\be
\ev{\imath\la_x^a}=-\nabla_{ix}A_{ix}^a
\ee
such that the only non-zero functional derivative of $\ev{\imath\la_x^a}$ is
\be
\ev{\imath A_{jy}^b\imath\la_x^a}=\imath\de^{ba}\nabla_{jx}\de(y-x).
\label{eq:la0}
\ee
All other functional derivatives of $\ev{\imath\la_x^a}$ vanish, even in 
the presence of sources.  Taking the functional derivative of \eq{eq:wtid2} 
with respect to $\imath\la_z^d$ gives
\bea
0&=&\int\dx{x}\de(t-x_0)\left\{
\frac{\imath}{g}\de(z-x)\ev{\imath\ov{c}_x^d}
-\frac{1}{g}\left[\frac{\nabla_{ix}}{(-\nabla_x^2)}
\ev{\imath\la_z^d\imath A_{ix}^a}\right]\ev{\imath\ov{c}_x^a}
\right.\nonumber\\&&\left.
-f^{abc}\ev{\imath\la_z^d\imath A_{ix}^a}t_{ij}(\vec{x})A_{jx}^bc_x^c
-f^{abc}\ev{\imath\la_z^d\imath A_{ix}^a}t_{ij}(\vec{x})
\ev{\imath J_{jx}^b\imath\ov{\et}_x^c}
\right.\nonumber\\&&\left.
-f^{abc}\ev{\imath\si_x^a}\frac{\de}{\de\imath\la_z^d}
\ev{\imath\ro_x^b\imath\ov{\et}_x^c}
-f^{abc}\ev{\imath A_{ix}^a}t_{ij}(\vec{x})\frac{\de}{\de\imath\la_z^d}
\ev{\imath J_{jx}^b\imath\ov{\et}_x^c}
+\frac{1}{2}f^{abc}\ev{\imath c_x^a}\frac{\de}{\de\imath\la_z^d}
\ev{\imath\ov{\et}_x^b\imath\ov{\et}_x^c}
\right\}.
\eea
Now, taking partial functional derivatives of the Legendre transform, 
\eq{eq:legtran}, one can show that
\bea
\frac{\de}{\de\imath\la_z^d}\ev{\imath J_\al\imath J_\ba}&\sim&
\ev{\imath J_\al\imath J_\ka}\ev{\imath\Phi_\ka\imath\la_z^d\imath\Phi_\ga}
\ev{\imath J_{\ga}\imath J_\ba}\nonumber\\
&=&0
\eea
since all three-point proper functions involving functional derivatives 
with respect to $\imath\la_z^d$ vanish.  Further using the result 
\eq{eq:la0} we arrive at
\be
0=\int\dx{x}\de(t-x_0)\left\{
\frac{\imath}{g}
\left[\de(z-x)-\frac{\nabla_{ix}\nabla_{iz}}{(-\nabla_x^2)}\de(x-z)\right]
\ev{\imath\ov{c}_x^d}
-\imath f^{dbc}\left[\nabla_{iz}\de(x-z)\right]t_{ij}(\vec{x})
\left[A_{jx}^bc_x^c+\ev{\imath J_{jx}^b\imath\ov{\et}_x^c}\right]
\right\}.
\ee
Knowing that $\nabla_{iz}\de(x-z)=-\nabla_{ix}\de(x-z)$ and using 
integration by parts on the last term, the above expression is a trivial 
identity.  Thus, even in the presence of sources, functional derivatives 
of \eq{eq:wtid2} with respect to $\imath\la_z^d$ give rise to an identity 
from which no further information can be obtained and we can set the 
classical field $\la_x^a=0$ (except for within partial derivatives used in 
conjunction with the Legendre transform) and ignore the Lagrange multiplier 
field from now on.

Equation (\ref{eq:wtid2}) is Grassmann valued and to proceed, we must first 
take the functional derivative with respect to $\imath c_z^d$.  Since we 
will be taking further derivatives, fields/sources must be retained (with 
the exception of $\la$ as discussed previously).  The subsequent \ST 
identities are thus functional derivatives of
\bea
0&=&\int\dx{x}\de(t-x_0)\left\{
-\frac{\imath}{g}\left[\pd_x^0\ev{\imath c_z^d\imath\si_x^a}\right]
\imath c_x^a
-\frac{\imath}{g}\left[\pd_x^0\ev{\imath\si_x^d}\right]\de(z-x)
\right.\nonumber\\&&
-\frac{1}{g}\left[\frac{\nabla_{ix}}{(-\nabla_x^2)}
\ev{\imath c_z^d\imath A_{ix}^a}\right]\ev{\imath\ov{c}_x^a}
+\frac{1}{g}\left[\frac{\nabla_{ix}}{(-\nabla_x^2)}
\ev{\imath A_{ix}^a}\right]\ev{\imath\ov{c}_x^a\imath c_z^d}
\nonumber\\&&
-f^{abc}\ev{\imath c_z^d\imath\si_x^a}
\left[\ev{\imath\ro_x^b\imath\ov{\et}_x^c}-\imath\si_x^b\imath c_x^c\right]
-f^{abc}\ev{\imath\si_x^a}
\left[\frac{\de}{\de\imath c_z^d}\ev{\imath\ro_x^b\imath\ov{\et}_x^c}
-\imath\si_x^b\de^{dc}\de(z-x)\right]
\nonumber\\&&
-f^{abc}\ev{\imath c_z^d\imath A_{ix}^a}t_{ij}(\vec{x})
\left[\ev{\imath J_{jx}^b\imath\ov{\et}_x^c}
-\imath A_{jx}^b\imath c_x^c\right]
-f^{abc}\ev{\imath A_{ix}^a}t_{ij}(\vec{x})
\left[\frac{\de}{\de\imath c_z^d}\ev{\imath J_{jx}^b\imath\ov{\et}_x^c}
-\imath A_{jx}^b\de^{dc}\de(z-x)\right]
\nonumber\\&&\left.
+\frac{1}{2}f^{abc}\ev{\imath c_z^d\imath c_x^a}
\left[\ev{\imath\ov{\et}_x^b\imath\ov{\et}_x^c}
-\imath c_x^b\imath c_x^c\right]
-\frac{1}{2}f^{abc}\ev{\imath c_x^a}
\left[\frac{\de}{\de\imath c_z^d}\ev{\imath\ov{\et}_x^b\imath\ov{\et}_x^c}
-2\de^{db}\de(z-x)\imath c_x^c\right]
\right\}.
\label{eq:wtid3}
\eea

\section{Slavnov--Taylor identities: two-point functions}
\setcounter{equation}{0}

The first of the \ST identities is generated by taking the functional 
derivative of \eq{eq:wtid3} with respect to $\imath\si_w^e$ and setting 
sources to zero.  The resulting equation reads
\bea
0&=&\int\dx{x}\de(t-x_0)\left\{
-\frac{\imath}{g}\left[\pd_x^0\ev{\imath\si_w^e\imath\si_x^d}\right]\de(z-x)
+\frac{1}{g}\left[\frac{\nabla_{ix}}{(-\nabla_x^2)}
\ev{\imath\si_w^e\imath A_{ix}^a}\right]\ev{\imath\ov{c}_x^a\imath c_z^d}
\right.\nonumber\\&&\left.
-f^{abc}\ev{\imath\si_w^e\imath\si_x^a}\left.\frac{\de}{\de\imath c_z^d}
\ev{\imath\ro_x^b\imath\ov{\et}_x^c}\right|_{J=0}
-f^{abc}\ev{\imath\si_w^e\imath A_{ix}^a}t_{ij}(\vec{x})
\left.\frac{\de}{\de\imath c_z^d}
\ev{\imath J_{jx}^b\imath\ov{\et}_x^c}\right|_{J=0}
\right\}.
\label{eq:xstid0}
\eea
This equation is best expressed in momentum space.  We define the momentum 
space two-point Green's functions via their respective Fourier transforms
\be
\ev{\imath\si_w^a\imath\si_x^b}=\int\dk{k}e^{-\imath k\cdot(w-x)}
\G_{\si\si}^{ab}(k_0,\vec{k}),\label{eq:ss0}
\ee
(similarly for $\G_{AA}$ etc. and also for the propagators $W_{AA}$ etc.) 
where the arguments of the momentum space functions reflect their 
noncovariant nature, we impose translational invariance in the usual way 
and $\dk{k}=d^4k/(2\pi)^4$.  We also define
\bea
f^{abc}\left.\frac{\de}{\de\imath c_z^d}
\ev{\imath\ro_x^b\imath\ov{\et}_x^c}\right|_{J=0}&=&
\int\dk{k}e^{-\imath k\cdot(x-z)}
\tilde{\Si}_{\si;\ov{c}c}^{ad}(k_0,\vec{k}),\label{eq:tscc0}\\
f^{abc}\left.\frac{\de}{\de\imath c_z^d}
\ev{\imath J_{jx}^b\imath\ov{\et}_x^c}\right|_{J=0}&=&
\int\dk{k}e^{-\imath k\cdot(x-z)}
\tilde{\Si}_{A;\ov{c}cj}^{ad}(k_0,\vec{k}).\label{eq:tacc0}
\eea
The latter two definitions and notations may at first sight appear somewhat 
artificial.  In fact, as will be shortly justified, their exact form in 
terms of standard Green's functions is irrelevant since neither term will 
contribute.  However, in the remainder of the paper, similar (but 
increasingly complicated) objects will occur so some explanation is in 
order.  The \DS equations are derived as functional derivatives of the 
field equations of motion and this gives rise to the familiar loop 
expressions.  For example (taken from Ref.~\cite{Watson:2007vc}),
\be
\frac{\de}{\de\imath A_{jw}^f}\int d^4yd^4z\,\G_{\ov{c}cAi}^{(0)bca}(y,z,x)
\ev{\imath\ov{\et}_z^c\imath\et_y^b}\;\longrightarrow\;
\int\dk{\w}\G_{\ov{c}cAi}^{(0)bca}(\w-k,-\w,k)W_{\ov{c}c}^{cd}(\w)
\G_{\ov{c}cAj}^{def}(\w,k-\w,-k)W_{\ov{c}c}^{eb}(\w-k)
\ee
gives the ghost loop term of the \DS equation for $\G_{AA}$ ($W_{\ov{c}c}$ 
is the ghost propagator, $\G_{\ov{c}cA}$ is the spatial ghost-gluon vertex 
and $\G_{\ov{c}cA}^{(0)}$ is the tree-level counterpart).  Importantly, 
all occurrences of objects such as $\ev{\imath\ov{\et}_z^c\imath\et_y^b}$ 
in the field equations of motion have an associated tree-level vertex -- 
these terms originate directly from the interaction terms of the original 
action such that the functional derivatives have a clear meaning as loops.  
However, in the case of the (nonabelian) Slavnov--Taylor identities one has 
a different structure which arises from the Gauss-BRST transform (or indeed 
generally from the BRS transform) and is exemplified in \eq{eq:wtid3}.  In 
this case, one has objects such as $\ev{\imath\ro_x^b\imath\ov{\et}_x^c}$ 
which must be functionally differentiated, but which are separated from any 
interaction type of factor.  However, even without the interaction term, 
these functional derivatives do still have a partial meaning as loop 
expressions; it is simply that the tree-level vertex is missing (although 
the color factor and momentum conservation are present).  We will denote 
such irregular expressions with a tilde (as in $\tilde{\Si}$ above or 
$\tilde{\G}$ later) and since these terms are important to the arguments 
presented in this study, we will write out their explicit forms where 
necessary.  As alluded to above, this type of pseudo-loop expression is 
common to the Slavnov--Taylor identities of nonabelian theories and in 
linear covariant gauges, there is a familiar example, referred to as the 
ghost-gluon scattering-like kernel which appears in the identity for the 
three-gluon vertex \cite{Marciano:1977su}.

Returning to \eq{eq:xstid0}, we can now write
\bea
0&=\int\dk{q_0}\dk{k}e^{-\imath q_0(t-z_0)-\imath k\cdot(w-z)}&\left\{
\frac{k_0}{g}\G_{\si\si}^{ed}(k_0,\vec{k})
-\frac{\imath}{g}\frac{k_i}{\vec{k}^2}\G_{\si Ai}^{ea}(k_0,\vec{k})
\G_{\ov{c}c}^{ad}(q_0+k_0,\vec{k})
\right.\nonumber\\&&\left.
-\G_{\si\si}^{ea}(k_0,\vec{k})
\tilde{\Si}_{\si;\ov{c}c}^{ad}(q_0+k_0,\vec{k})
-\G_{\si Ai}^{ea}(k_0,\vec{k})t_{ij}(\vec{k})
\tilde{\Si}_{A;\ov{c}cj}^{ad}(k_0+q_0,\vec{k})
\right\}.
\eea
In the above, notice how the $\de(t-x_0)$ factor characterizing the 
Gauss-BRST transform leads to the energy insertion $q_0$ which breaks the 
energy flow through the various Green's functions and in particular how 
this insertion affects only the ghost functions.  Since $t$, $w$ and $z$ 
are arbitrary, we can write down the momentum space \ST identity:
\be
\frac{k_0}{g}\G_{\si\si}^{ed}(k_0,\vec{k})=
\frac{\imath}{g}\frac{k_i}{\vec{k}^2}\G_{\si Ai}^{ea}(k_0,\vec{k})
\G_{\ov{c}c}^{ad}(q_0+k_0,\vec{k})
+\G_{\si\si}^{ea}(k_0,\vec{k})
\tilde{\Si}_{\si;\ov{c}c}^{ad}(q_0+k_0,\vec{k})
+\G_{\si Ai}^{ea}(k_0,\vec{k})t_{ij}(\vec{k})
\tilde{\Si}_{A;\ov{c}cj}^{ad}(k_0+q_0,\vec{k}).
\label{eq:mstid0}
\ee
Now consider the definition of $\tilde{\Si}_{Aj;\ov{c}c}$, given by 
\eq{eq:tacc0}: under a parity transform, the vector source $J_{jx}^b$ 
changes sign \cite{Watson:2006yq} and we see that
\be
\tilde{\Si}_{A;\ov{c}cj}^{ad}(k_0,-\vec{k})=
-\tilde{\Si}_{A;\ov{c}cj}^{ad}(k_0,\vec{k})
\ee
from which we can infer that
\be
t_{ij}(\vec{k})\tilde{\Si}_{A;\ov{c}cj}^{ad}(k_0+q_0,\vec{k})
\sim t_{ij}(\vec{k})k_j=0
\label{eq:tilde0}
\ee
such that the last term of \eq{eq:mstid0} vanishes.  Because the energy 
scale, $q_0$, is arbitrary, we can make a further translation 
$q_0\rightarrow-q_0-2k_0$ and \eq{eq:mstid0} becomes
\be
\frac{k_0}{g}\G_{\si\si}^{ed}(k_0,\vec{k})=
\frac{\imath}{g}\frac{k_i}{\vec{k}^2}\G_{\si Ai}^{ea}(k_0,\vec{k})
\G_{\ov{c}c}^{ad}(-q_0-k_0,\vec{k})
+\G_{\si\si}^{ea}(k_0,\vec{k})
\tilde{\Si}_{\si;\ov{c}c}^{ad}(-q_0-k_0,\vec{k}).
\label{eq:mstid1}
\ee
Knowing that the $\si$-field and $\ro$-source change sign under 
time-reversal, whereas the ghost-antighost pair is invariant 
\cite{Watson:2006yq}, we have that
\bea
\tilde{\Si}_{\si;\ov{c}c}^{ad}(-q_0-k_0,\vec{k})&=&
-\tilde{\Si}_{\si;\ov{c}c}^{ad}(q_0+k_0,\vec{k}),\nonumber\\
\G_{\ov{c}c}^{ad}(-q_0-k_0,\vec{k})&=&\G_{\ov{c}c}^{ad}(q_0+k_0,\vec{k}),
\eea
such that
\be
\frac{k_0}{g}\G_{\si\si}^{ed}(k_0,\vec{k})=
\frac{\imath}{g}\frac{k_i}{\vec{k}^2}\G_{\si Ai}^{ea}(k_0,\vec{k})
\G_{\ov{c}c}^{ad}(q_0+k_0,\vec{k})
-\G_{\si\si}^{ea}(k_0,\vec{k})\tilde{\Si}_{\si;\ov{c}c}^{ad}(q_0+k_0,\vec{k}).
\label{eq:mstid2}
\ee
Comparing this with the original \eq{eq:mstid0} above, we find that
\be
\tilde{\Si}_{\si;\ov{c}c}^{ad}(q_0+k_0,\vec{k})=0.
\label{eq:tilde1}
\ee
The \ST identity now reads
\be
\frac{k_0}{g}\G_{\si\si}^{ed}(k_0,\vec{k})=
\frac{\imath}{g}\frac{k_i}{\vec{k}^2}\G_{\si Ai}^{ea}(k_0,\vec{k})
\G_{\ov{c}c}^{ad}(q_0+k_0,\vec{k}).
\ee
However, since $q_0$ is arbitrary, we can further state that the two-point 
ghost proper function $\G_{\ov{c}c}$ is independent of energy, i.e.,
\be
\G_{\ov{c}c}^{ad}(q_0+k_0,\vec{k})\rightarrow\G_{\ov{c}c}^{ad}(\vec{k}).
\ee
This was known to all orders in perturbation theory 
\cite{Watson:2006yq,Watson:2007vc}, but it is reassuring that the result is 
confirmed nonperturbatively.  We now have the final form of the first of 
the \ST identities:
\be
\frac{k_0}{g}\G_{\si\si}^{ed}(k_0,\vec{k})=
\frac{\imath}{g}\frac{k_i}{\vec{k}^2}\G_{\si Ai}^{ea}(k_0,\vec{k})
\G_{\ov{c}c}^{ad}(\vec{k}).
\label{eq:stid0}
\ee

Repeating the above analysis but starting by functionally differentiating 
\eq{eq:wtid3} with respect to $\imath A_{kw}^e$ and setting sources to zero 
leads to the similar \ST identity
\be
\frac{k_0}{g}\G_{A\si k}^{ed}(k_0,\vec{k})=
\frac{\imath}{g}\frac{k_i}{\vec{k}^2}\G_{AAki}^{ea}(k_0,\vec{k})
\G_{\ov{c}c}^{ad}(\vec{k})
\label{eq:stid1}.
\ee

Now, since the energy, $k_0$, is a scalar quantity we see immediately from 
Eqs.~(\ref{eq:stid0}) and (\ref{eq:stid1}) that the two-point proper 
Green's functions involving functional derivatives with respect to the 
temporal gluon field, $\si$, (these will be referred to as the temporal 
Green's functions) can be unambiguously expressed in terms of the ghost 
and (longitudinal) spatial gluon two-point proper Green's functions.  (We 
will discuss the identities, Eqs.~(\ref{eq:stid0}) and (\ref{eq:stid1}), 
and their general kinematical decompositions further in 
Section~\ref{sec:five}.)  In a sense, up to the level of the two-point 
functions the field $\si$ has been eliminated from the system or 
equivalently has been integrated out of the functional form of the 
action.  The two identities, Eqs.~(\ref{eq:stid0}) and (\ref{eq:stid1}), 
have been derived previously from the standard BRS invariance, using 
perturbative arguments to eliminate the $\tilde{\Si}$ terms and were 
verified to one-loop order in perturbation theory \cite{Watson:2007vc}.  
The above derivation is however slightly superior since it makes reference 
only to symmetry arguments and does not rely on the implicit assumption 
that the all-orders resummation of perturbation theory is equivalent to the 
nonperturbative theory.  Together, the two identities are the Coulomb gauge 
equivalent of the standard covariant gauge result that the longitudinal part 
of the gluon polarization is not dressed 
\cite{Taylor:1971ff,Slavnov:1972fg}.  This also highlights an important 
difference between Landau gauge and Coulomb gauge: in the former, the gluon 
polarization is transverse to the four-momentum; in the latter, the gluon 
polarization ($\G_{AA}$ in our notation) is explicitly \emph{not} 
transverse --- instead, the spatially longitudinal, ghost and temporal 
two-point proper functions are all related.  This connection is reminiscent 
of the quartet mechanism in the Kugo--Ojima confinement scenario 
\cite{Kugo:1979gm}.  This is hardly surprising since the Kugo--Ojima 
confinement scenario is based on the identification of a well-defined 
conserved color charge and the Slavnov--Taylor identities here are the 
dynamical expression of charge conservation.

\section{Slavnov--Taylor identities: vertex functions}
\setcounter{equation}{0}

In this section, we study further functional derivatives of \eq{eq:wtid3} 
and show that the resultant Slavnov--Taylor identities eventually form a 
closed set from which the temporal Green's functions (i.e., those with at 
least one external $\si$-leg) can be unambiguously derived.  This closed set 
turns out to be somewhat extended and so, for clarity of presentation, we 
separate the various classes of functional derivatives according to how many 
pairs of ghost/antighost functional derivatives are present.  By explicitly 
separating the Grassmann-valued fields, we may then modify our notation for 
the generic field $\Phi_\al$ or source $J_\al$ to be restricted to only the 
$\vec{A}$ or $\si$ fields/sources and their associated attributes which 
allows us to compactify the formalism.

\subsection{No further ghost derivatives}
Let us begin by functionally differentiating \eq{eq:wtid3} twice with 
respect to $\imath\Phi_{\la(k)w}^e$ and $\imath\Phi_{\ta(l)v}^f$ where as 
mentioned above, the indices $\la(k)$ and $\ta(l)$ refer here to either the 
$A$-field with its associated spatial index ($k$ or $l$) or to the 
$\si$-field with no associated spatial index.  Setting sources to zero, the 
resulting equation reads:
\bea
0&=&\int\dx{x}\de(t-x_0)\left\{
-\frac{\imath}{g}\left[\pd_x^0
\ev{\imath\Phi_{\ta(l)v}^f\imath\Phi_{\la(k)w}^e\imath\si_x^d}\right]\de(z-x)
+\frac{1}{g}\left[\frac{\nabla_{ix}}{(-\nabla_x^2)}
\ev{\imath\Phi_{\ta(l)v}^f\imath\Phi_{\la(k)w}^e\imath A_{ix}^a}\right]
\ev{\imath\ov{c}_x^a\imath c_z^d}
\right.\nonumber\\&&
+\frac{1}{g}\left[\frac{\nabla_{ix}}{(-\nabla_x^2)}
\ev{\imath\Phi_{\la(k)w}^e\imath A_{ix}^a}\right]
\ev{\imath\ov{c}_x^a\imath c_z^d\imath\Phi_{\ta(l)v}^f}
+\frac{1}{g}\left[\frac{\nabla_{ix}}{(-\nabla_x^2)}
\ev{\imath\Phi_{\ta(l)v}^f\imath A_{ix}^a}\right]
\ev{\imath\ov{c}_x^a\imath c_z^d\imath\Phi_{\la(k)w}^e}
\nonumber\\&&
-\ev{\imath\Phi_{\la(k)w}^e\imath\si_x^a}
\left[f^{abc}\left.\frac{\de^2}{\de\imath\Phi_{\ta(l)v}^f\de\imath c_z^d}
\ev{\imath\ro_x^b\imath\ov{\et}_x^c}\right|_{J=0}
-f^{afd}\de_{\si\ta}\de(v-x)\de(z-x)\right]
\nonumber\\&&
-\ev{\imath\Phi_{\ta(l)v}^f\imath\si_x^a}
\left[f^{abc}\left.\frac{\de^2}{\de\imath\Phi_{\la(k)w}^f\de\imath c_z^d}
\ev{\imath\ro_x^b\imath\ov{\et}_x^c}\right|_{J=0}
-f^{aed}\de_{\si\la}\de(w-x)\de(z-x)\right]
\nonumber\\&&
-\ev{\imath\Phi_{\la(k)w}^e\imath A_{ix}^a}t_{ij}(\vec{x})
\left[f^{abc}\left.\frac{\de^2}{\de\imath\Phi_{\ta(l)v}^f\de\imath c_z^d}
\ev{\imath J_{jx}^b\imath\ov{\et}_x^c}\right|_{J=0}
-f^{afd}\de_{jl}\de_{A\ta}\de(v-x)\de(z-x)\right]
\nonumber\\&&\left.
-\ev{\imath\Phi_{\ta(l)v}^f\imath A_{ix}^a}t_{ij}(\vec{x})
\left[f^{abc}\left.\frac{\de^2}{\de\imath\Phi_{\la(k)w}^e\de\imath c_z^d}
\ev{\imath J_{jx}^b\imath\ov{\et}_x^c}\right|_{J=0}
-f^{aed}\de_{jk}\de_{A\la}\de(w-x)\de(z-x)\right]
\right\}.
\eea
In the above, we have made use of the results Eqs.~(\ref{eq:tilde0}) and 
(\ref{eq:tilde1}) to eliminate such terms.  Because the indices $\la$ and 
$\ta$ refer to field types that commute, the above equation actually 
represents three separate equations.  Let us define
\bea
\tilde{\G}_{\si;\ov{c}c\ta(l)}^{adf}(x,z,v)&=&
gf^{abc}\left.\frac{\de^2}{\de\imath\Phi_{\ta(l)v}^f\de\imath c_z^d}
\ev{\imath\ro_x^b\imath\ov{\et}_x^c}\right|_{J=0},
\nonumber\\
\tilde{\G}_{A;\ov{c}c\ta j(l)}^{adf}(x,z,v)&=&
gf^{abc}\left.\frac{\de^2}{\de\imath\Phi_{\ta(l)v}^f\de\imath c_z^d}
\ev{\imath J_{jx}^b\imath\ov{\et}_x^c}\right|_{J=0},
\label{eq:tildedef}
\eea
(these expressions will be explained later).  We further notice that by 
taking two functional derivatives of \eq{eq:gheom0} and setting sources to 
zero then we have
\be
\ev{\imath\ov{c}_x^a\imath c_z^d\imath\Phi_{\ta(l)v}^f}=
-gf^{adf}\de_{A\ta}\de_{jl}\nabla_{jx}\de(v-x)\de(z-x)
-\nabla_{jx}\tilde{\G}_{A;\ov{c}c\ta j(l)}^{adf}(x,z,v).
\label{eq:ghgldse0}
\ee
This equation is of course the precursor to the \DS equation for the 
temporal and spatial ghost-gluon vertices.  Our identities can thus be 
written
\bea
0&=&\int\dx{x}\de(t-x_0)\left\{
-\imath\left[\pd_x^0
\ev{\imath\Phi_{\ta(l)v}^f\imath\Phi_{\la(k)w}^e\imath\si_x^d}\right]\de(z-x)
+\left[\frac{\nabla_{ix}}{(-\nabla_x^2)}
\ev{\imath\Phi_{\ta(l)v}^f\imath\Phi_{\la(k)w}^e\imath A_{ix}^a}\right]
\ev{\imath\ov{c}_x^a\imath c_z^d}
\right.\nonumber\\&&
-\ev{\imath\Phi_{\la(k)w}^e\imath A_{ix}^a}
\left[\tilde{\G}_{A;\ov{c}c\ta i(l)}^{adf}(x,z,v)
+gf^{adf}\de_{il}\de_{A\ta}\de(v-x)\de(z-x)\right]
\nonumber\\&&
-\ev{\imath\Phi_{\ta(l)v}^f\imath A_{ix}^a}
\left[\tilde{\G}_{A;\ov{c}c\la i(k)}^{ade}(x,z,w)
+gf^{ade}\de_{ik}\de_{A\la}\de(w-x)\de(z-x)\right]
\nonumber\\&&
-\ev{\imath\Phi_{\la(k)w}^e\imath\si_x^a}
\left[\tilde{\G}_{\si;\ov{c}c\ta(l)}^{adf}(x,z,v)
+gf^{adf}\de_{\si\ta}\de(v-x)\de(z-x)\right]
\nonumber\\&&\left.
-\ev{\imath\Phi_{\ta(l)v}^f\imath\si_x^a}
\left[\tilde{\G}_{\si;\ov{c}c\la(k)}^{ade}(x,z,w)
+gf^{ade}\de_{\si\la}\de(w-x)\de(z-x)\right]
\right\}.
\label{eq:inter0}
\eea
Let us now introduce our convention for the Fourier transform of the 
three-point functions (similarly for higher $n$-point functions):
\be
f(x,y,z)=\int\dk{k_1}\dk{k_2}\dk{k_3}(2\pi)^4\de(k_1+k_2+k_3)
e^{-\imath k_1\cdot x-\imath k_2\cdot y-\imath k_3\cdot z}f(k_1,k_2,k_3).
\ee
All momenta of the three-point functions are defined as incoming and for 
brevity we do not split the temporal and spatial sets of arguments.  
Returning to \eq{eq:inter0}, after some manipulation we arrive at the 
three \ST identities ($k_1+k_2+k_3=0$):
\bea
k_3^0\G_{\ta\la\si(lk)}^{fed}(k_1,k_2,k_3)
&=&
\imath\frac{k_{3i}}{\vec{k}_3^2}\G_{\ta\la A(lk)i}^{fea}(k_1,k_2,k_3)
\G_{\ov{c}c}^{ad}(-\vec{k}_3)
\nonumber\\&&
-\G_{\la A(k)i}^{ea}(k_2)
\left[\tilde{\G}_{A;\ov{c}c\ta i(l)}^{adf}(k_2+q_0,k_3-q_0,k_1)
+gf^{adf}\de_{il}\de_{A\ta}\right]
\nonumber\\&&
-\G_{\ta A(l)i}^{fa}(k_1)
\left[\tilde{\G}_{A;\ov{c}c\la i(k)}^{ade}(k_1+q_0,k_3-q_0,k_2)
+gf^{ade}\de_{ik}\de_{A\la}\right]
\nonumber\\&&
-\G_{\la\si(k)}^{ea}(k_2)
\left[\tilde{\G}_{\si;\ov{c}c\ta(l)}^{adf}(k_2+q_0,k_3-q_0,k_1)
+gf^{adf}\de_{\si\ta}\right]
\nonumber\\&&
-\G_{\ta\si(l)}^{fa}(k_1)
\left[\tilde{\G}_{\si;\ov{c}c\la(k)}^{ade}(k_1+q_0,k_3-q_0,k_2)
+gf^{ade}\de_{\si\la}\right].
\label{eq:stid2}
\eea
These identities can easily be verified at tree-level, using the Feynman 
rules presented in Ref.~\cite{Watson:2007vc} (and for completeness presented 
in Appendix~\ref{app:decomp}).  They are the Coulomb gauge analogue of the 
familiar Slavnov--Taylor identity for the three-gluon vertex 
\cite{Marciano:1977su}.  As in the previous section, the energy $q_0$ is 
injected into the ghost line of the various factors but one cannot cancel 
terms using symmetry properties anymore (as was the case for the 
$\tilde{\Sigma}$ kernels) because of the presence of other energy arguments 
in the equations.  Neglecting the $\tilde{\G}$ terms, we can see that each 
vertex involving a functional derivative with respect to the temporal 
$\si$-field is fully determined (again because $k_3^0$ is a scalar quantity) 
given the corresponding (spatially contracted) vertex involving the 
derivative with respect to the spatial $A$-field.

Let us now discuss the form of the $\tilde{\G}$ factors.  From the 
definition, \eq{eq:tildedef}, and taking functional derivatives in standard 
fashion, we have
\bea
\tilde{\G}_{\si;\ov{c}c\ta}^{adf}(x,z,v)&=&
gf^{abc}\left.\frac{\de^2}{\de\imath\Phi_{\ta v}^f\de\imath c_z^d}
\ev{\imath\ro_x^b\imath\ov{\et}_x^c}\right|_{J=0}\nonumber\\
&=&gf^{abc}\left\{
\ev{\imath\ro_x^b\imath J_\e}
\ev{\imath\Phi_\e\imath\Phi_{\ta v}^f\imath\Phi_\la}
\ev{\imath J_\la\imath J_\ka}\ev{\imath\ov{\et}_x^c\imath\et_u^g}
\ev{\imath\ov{c}_u^g\imath c_z^d\imath\Phi_\ka}
\right.\nonumber\\&&
-\ev{\imath\ro_x^b\imath J_\ka}\ev{\imath\ov{\et}_x^c\imath\et_u^g}
\ev{\ov{c}_u^g\imath c_z^d\imath\Phi_\ka\imath\Phi_{\ta(l)v}^f}
\nonumber\\&&\left.
+\ev{\imath\ro_x^b\imath J_\ka}\ev{\imath\ov{\et}_x^c\imath\et_u^g}
\ev{\imath\ov{c}_u^g\imath c_s^h\imath\Phi_{\ta v}^f}
\ev{\imath\ov{\et}_s^h\et_r^i}\ev{\imath\ov{c}_r^i\imath c_z^d\imath\Phi_\ka}
\right\}
\label{eq:rnd1}
\eea
where we recall that having explicitly extracted the Grassmann-valued 
fields, the repeated indices refer here to only the $\si$ or $A$-fields.  
The coordinates $u$, $s$ and $r$ are implicitly integrated over.  We further 
omit the possible spatial index ($l$) associated when $\ta$ refers to the 
$\vec{A}$-field for notational convenience.  A similar expression exists 
for $\tilde{\G}_{A;\ov{c}c\ta}$.  In momentum space, we have ($p_1+p_2+p_3=0$)
\bea
\tilde{\G}_{\si;\ov{c}c\ta}^{adf}(p_1,p_2,p_3)&=&
gf^{abc}\int\dk{k}W_{\si\e}^{bg}(k)\G_{\e\ta\mu}^{gfh}(k,p_3,-k-p_3)
W_{\mu\ka}^{hi}(k+p_3)W_{\ov{c}c}^{cj}(p_1-k)
\G_{\ov{c}c\ka}^{jdi}(p_1-k,p_2,k+p_3)
\nonumber\\&&
-gf^{abc}\int\dk{k}W_{\si\ka}^{bg}(k)W_{\ov{c}c}^{ch}(p_1-k)
\G_{\ov{c}c\ka\ta}^{hdgf}(p_1-k,p_2,k,p_3)
\nonumber\\&&
+gf^{abc}\int\dk{k}W_{\si\ka}^{bg}(k)W_{\ov{c}c}^{ch}(p_1-k)
\G_{\ov{c}c\ta}^{hif}(p_1-k,p_2+k,p_3)W_{\ov{c}c}^{ij}(-p_2-k)
\G_{\ov{c}c\ka}^{jdg}(-p_2-k,p_2,k),
\nonumber\\&&
\label{eq:ghgldse1}
\eea
(we make the internal color indices explicit) again with similar expressions 
for the other $\tilde{\G}$ factors.  Given that there is the implicit 
summation over $\e$, $\mu$ and $\ka$ (referring to the field types $\si$ 
and $A$), what this means is that the $\tilde{\G}$ considered so far are 
in general dependent not only on the non-ghost temporal three-point 
functions that are explicit in \eq{eq:stid2} (thus showing that 
\eq{eq:stid2} actually forms a set of nonlinear integral equations) but on 
the further temporal Green's functions $\G_{\ov{c}c\si}$, 
$\G_{\ov{c}c\si A}$ and $\G_{\ov{c}c\si\si}$ (and of course all the other 
two-point functions, the purely spatial $A$ and the ghost three- and 
four-point functions).

\subsection{One pair of further ghost derivatives}
Let us now consider the \ST identities that arise if we functionally 
differentiate \eq{eq:wtid3} with respect to at least $\imath\ov{c}_v^f$ and 
$\imath c_w^e$.  If we restrict ourselves to considering no further 
ghost/antighost functional derivatives, then we are able to set the 
corresponding fields/sources to zero whilst maintaining the rest without 
confusion.  The resulting expression reads:
\bea
0&=&\int\dx{x}\de(t-x_0)\left\{
\frac{\imath}{g}\left[\pd_x^0
\ev{\imath\ov{c}_v^f\imath c_z^d\imath\si_x^e}\right]\de(w-x)
-\frac{\imath}{g}\left[\pd_x^0
\ev{\imath\ov{c}_v^f\imath c_w^e\imath\si_x^d}\right]\de(z-x)
\right.\nonumber\\&&
-\frac{1}{g}\left[\frac{\nabla_{ix}}{(-\nabla_x^2)}
\ev{\imath\ov{c}_v^f\imath c_z^d\imath A_{ix}^a}\right]
\ev{\imath\ov{c}_x^a\imath c_w^e}
+\frac{1}{g}\left[\frac{\nabla_{ix}}{(-\nabla_x^2)}
\ev{\imath\ov{c}_v^f\imath c_w^e\imath A_{ix}^a}\right]
\ev{\imath\ov{c}_x^a\imath c_z^d}
\nonumber\\&&
+\frac{1}{g}\left[\frac{\nabla_{ix}}{(-\nabla_x^2)}
\ev{\imath A_{ix}^a}\right]
\ev{\imath\ov{c}_v^f\imath c_w^e\imath\ov{c}_x^a\imath c_z^d}
\nonumber\\&&
+f^{abc}\ev{\imath\ov{c}_v^f\imath c_z^d\imath\si_x^a}
\left[\frac{\de}{\de\imath c_w^e}
\ev{\imath\ro_x^b\imath\ov{\et}_x^c}-\imath\si_x^b\de^{ec}\de(w-x)\right]
-f^{abc}\ev{\imath\ov{c}_v^f\imath c_w^e\imath\si_x^a}
\left[\frac{\de}{\de\imath c_z^d}
\ev{\imath\ro_x^b\imath\ov{\et}_x^c}-\imath\si_x^b\de^{dc}\de(z-x)\right]
\nonumber\\&&
-f^{abc}\ev{\imath\si_x^a}
\frac{\de^3}{\de\imath\ov{c}_v^f\de\imath c_w^e\de\imath c_z^d}
\ev{\imath\ro_x^b\imath\ov{\et}_x^c}
+f^{abc}\ev{\imath\ov{c}_v^f\imath c_z^d\imath A_{ix}^a}t_{ij}(\vec{x})
\left[\frac{\de}{\de\imath c_w^e}
\ev{\imath J_{jx}^b\imath\ov{\et}_x^c}-\imath A_{jx}^b\de^{ec}\de(w-x)\right]
\nonumber\\&&
-f^{abc}\ev{\imath\ov{c}_v^f\imath c_w^e\imath A_{ix}^a}t_{ij}(\vec{x})
\left[\frac{\de}{\de\imath c_z^d}
\ev{\imath J_{jx}^b\imath\ov{\et}_x^c}-\imath A_{jx}^b\de^{dc}\de(z-x)\right]
-f^{abc}\ev{\imath A_{ix}^a}t_{ij}(\vec{x})
\frac{\de^3}{\de\imath\ov{c}_v^f\de\imath c_w^e\de\imath c_z^d}
\ev{\imath J_{jx}^b\imath\ov{\et}_x^c}
\nonumber\\&&\left.
+\frac{1}{2}f^{abc}\ev{\ov{c}_v^f\imath c_x^a}
\left[\frac{\de^2}{\de\imath c_w^e\de\imath c_z^d}
\ev{\imath\ov{\et}_x^b\imath\ov{\et}_x^c}
-2\de^{db}\de^{ec}\de(z-x)\de(w-x)\right]
\right\}_{\ov{\et}=\et=0}.
\label{eq:wtid4}
\eea
To derive the identity for the ghost-gluon vertex, we further set the 
remaining sources to zero (we will return to \eq{eq:wtid4} later in this 
subsection to derive more identities) and again using the results given by 
Eqs.~(\ref{eq:tilde0}) and (\ref{eq:tilde1}) we get
\bea
0&=&\int\dx{x}\de(t-x_0)\left\{
\frac{\imath}{g}
\left[\pd_x^0\ev{\imath\ov{c}_v^f\imath c_z^d\imath\si_x^e}\right]\de(w-x)
-\frac{\imath}{g}
\left[\pd_x^0\ev{\imath\ov{c}_v^f\imath c_w^e\imath\si_x^d}\right]\de(z-x)
\right.\nonumber\\&&
-\frac{1}{g}\left[\frac{\nabla_{ix}}{(-\nabla_x^2)}
\ev{\imath\ov{c}_v^f\imath c_z^d\imath A_{ix}^a}\right]
\ev{\imath\ov{c}_x^a\imath c_w^e}
+\frac{1}{g}\left[\frac{\nabla_{ix}}{(-\nabla_x^2)}
\ev{\imath\ov{c}_v^f\imath c_w^e\imath A_{ix}^a}\right]
\ev{\imath\ov{c}_x^a\imath c_z^d}
\nonumber\\&&\left.
+\frac{1}{2g}\ev{\imath\ov{c}_v^f\imath c_x^a}
\tilde{\G}_{\ov{c};\ov{c}cc}^{ade}(x,z,w)
-f^{ade}\ev{\imath\ov{c}_v^f\imath c_x^a}\de(z-x)\de(w-x)
\right\}.
\eea
In the above, we have defined
\bea
\tilde{\G}_{\ov{c};\ov{c}cc}^{ade}(x,z,w)&=&
gf^{abc}\left.\frac{\de^2}{\de\imath c_w^e\de\imath c_z^d}
\ev{\imath\ov{\et}_x^b\imath\ov{\et}_x^c}\right|_{J=0}
\nonumber\\
&=&gf^{abc}\ev{\imath\ov{\et}_x^b\imath\et_\nu}
\ev{\imath\ov{\et}_x^c\imath\et_\mu}
\left[2\ev{\imath J_\ka\imath J_\e}
\ev{\imath\ov{c}_\nu\imath c_z^d\imath\Phi_\ka}
\ev{\imath\ov{c}_\mu\imath c_w^e\imath\Phi_\e}
-\ev{\imath\ov{c}_\nu\imath c_w^e\imath c_z^d\imath\ov{c}_\mu}\right]
\eea
(internal indices $\ka$ and $\e$ referring only to $A$ or $\si$-fields as 
in the previous subsection).  Considering now the Fourier transform and 
using the same conventions as before, we arrive at the \ST identity for the 
temporal ghost-gluon vertex
\bea
0&=&k_3^0\G_{\ov{c}c\si}^{fde}(k_1,k_2,k_3)
-\frac{\imath k_{3i}}{\vec{k}_3^2}\G_{\ov{c}cAi}^{fda}(k_1,k_2,k_3)
\G_{\ov{c}c}^{ae}(-\vec{k}_3)
\nonumber\\&&
-(k_2^0+q_0)\G_{\ov{c}c\si}^{fed}(k_1,k_3-q_0,k_2+q_0)
+\frac{\imath k_{2i}}{\vec{k}_2^2}\G_{\ov{c}cAi}^{fea}(k_1,k_3-q_0,k_2+q_0)
\G_{\ov{c}c}^{ad}(-\vec{k}_2)
\nonumber\\&&
+\frac{1}{2}\G_{\ov{c}c}^{fa}(\vec{k}_1)
\tilde{\G}_{\ov{c};\ov{c}cc}^{ade}(k_1+q_0,k_2,k_3-q_0)
-gf^{ade}\G_{\ov{c}c}^{fa}(\vec{k}_1)
\label{eq:stid3}
\eea
with
\bea
\tilde{\G}_{\ov{c};\ov{c}cc}^{ade}(p_1,p_2,p_3)&=
gf^{abc}\int\dk{k}W_{\ov{c}c}^{bf}(p_1-k)W_{\ov{c}c}^{cg}(k)&
\left\{2\G_{\ov{c}c\ka}^{fdh}(p_1-k,p_2,p_3+k)
\G_{\ov{c}c\e}^{gei}(k,p_3,-p_3-k)W_{\ka\e}^{hi}(-p_3-k)
\right.\nonumber\\&&\left.
-\G_{\ov{c}cc\ov{c}}^{fedg}(p_1-k,p_3,p_2,k)
\right\}.
\eea
Recall that we are free to choose the energy injection, $q_0$, at will.  If 
we set $k_2^0+q_0=0$ in \eq{eq:stid3}, then (given that for such general 
spacelike momentum configurations, the Green's functions can have no 
singularities) the $\G_{\ov{c}c\si}$ term of the middle line drops out and 
we have an unambiguous nonlinear integral equation for $\G_{\ov{c}c\si}$.  
Crucially, as far as the three-point functions involving the temporal 
$\si$-field are concerned, the equation forms a closed expression without 
approximation.  One can immediately verify this identity at tree-level using 
the Feynman rules of Ref.~\cite{Watson:2007vc} (and summarized in 
Appendix~\ref{app:decomp}).  A similar Slavnov--Taylor identity for the 
ghost-gluon vertex in Landau gauge does in fact exist 
\cite{von Smekal:1997vx,Alkofer:2000wg}.

Returning to \eq{eq:wtid4}, we take the further functional derivative with 
respect to $\imath\Phi_{\la(k)u}^g$ (again the field type $\la$ refers only 
to either the spatial $A$-field with index $k$ or to the temporal 
$\si$-field) and set sources to zero.  Making use of the results 
Eqs.~(\ref{eq:tilde0}) and (\ref{eq:tilde1}) to eliminate such terms and if 
we further use the appropriate functional derivatives of \eq{eq:gheom0} as 
before, then we have \eq{eq:inter1} (such lengthy configuration space 
expressions are relegated to Appendix~\ref{app:eqlist}).  
Equation~(\ref{eq:inter1}) involves three new kernels: 
$\tilde{\G}_{\si;\ov{c}cc\ov{c}}$ and $\tilde{\G}_{\ov{c};\ov{c}cc\la}$ are 
given by Eqs.~(\ref{eq:rnd2}) and (\ref{eq:rnd3}), respectively; the third 
kernel, $\tilde{\G}_{A;\ov{c}cc\ov{c}}$, has a similar expression to 
\eq{eq:rnd2}.  There is only one possible new Green's function involving 
derivatives with respect to the $\si$-field and this is 
$\G_{\ov{c}c\ov{c}c\si}$.  Taking the Fourier transform of \eq{eq:inter1} 
leads to the following \ST identity in momentum space ($\sum k_a=0$):
\bea
0&=&k_4^0\G_{\ov{c}c\la\si(k)}^{fdge}(k_1,k_2,k_3,k_4)
-\frac{\imath k_{4i}}{\vec{k}_4^2}
\G_{\ov{c}c\la A(k)i}^{fdga}(k_1,k_2,k_3,k_4)\G_{\ov{c}c}^{ae}(-\vec{k}_4)
\nonumber\\&&
-(k_2^0+q_0)\G_{\ov{c}c\la\si(k)}^{fegd}(k_1,k_4-q_0,k_3,k_2+q_0)
+\frac{\imath k_{2i}}{\vec{k}_2^2}
\G_{\ov{c}c\la A(k)i}^{fega}(k_1,k_4-q_0,k_3,k_2+q_0)
\G_{\ov{c}c}^{ad}(-\vec{k}_2)
\nonumber\\&&
+\G_{\ov{c}c\si}^{fda}(k_1,k_2,k_3+k_4)
\left[\tilde{\G}_{\si;\ov{c}c\la(k)}^{aeg}(q_0-k_3-k_4,k_4-q_0,k_3)
-gf^{age}\de_{\si\la}\right]
\nonumber\\&&
+\G_{\ov{c}cAi}^{fda}(k_1,k_2,k_3+k_4)
\left[\tilde{\G}_{A;\ov{c}c\la i(k)}^{aeg}(q_0-k_3-k_4,k_4-q_0,k_3)
-gf^{age}\de_{ki}\de_{A\la}\right]
\nonumber\\&&
-\G_{\ov{c}c\si}^{fea}(k_1,k_4-q_0,k_2+k_3+q_0)
\left[\tilde{\G}_{\si;\ov{c}c\la(k)}^{adg}(-k_2-k_3,k_2,k_3)
-gf^{agd}\de_{\si\la}\right]
\nonumber\\&&
-\G_{\ov{c}cAi}^{fea}(k_1,k_4-q_0,k_2+k_3+q_0)
\left[\tilde{\G}_{A;\ov{c}c\la i(k)}^{adg}(-k_2-k_3,k_2,k_3)
-gf^{agd}\de_{ki}\de_{A\la}\right]
\nonumber\\&&
-\G_{\la\si(k)}^{ga}(k_3)
\tilde{\G}_{\si;\ov{c}cc\ov{c}}^{adef}(k_3+q_0,k_2,k_4-q_0,k_1)
-\G_{\la A(k)i}^{ga}(k_3)
\tilde{\G}_{A;\ov{c}cc\ov{c}i}^{adef}(k_3+q_0,k_2,k_4-q_0,k_1)
\nonumber\\&&
+\frac{1}{2}\G_{\ov{c}c\la(k)}^{fag}(k_1,k_2+k_4,k_3)
\left[\tilde{\G}_{\ov{c};\ov{c}cc}^{ade}(q_0-k_2-k_4,k_2,k_4-q_0)
-2gf^{ade}\right]
\nonumber\\&&
+\frac{1}{2}\G_{\ov{c}c}^{fa}(\vec{k}_1)
\tilde{\G}_{\ov{c};\ov{c}cc\la(k)}^{adeg}(k_1+q_0,k_2,k_4-q_0,k_3),
\label{eq:stid4}
\eea
with the kernels (because of the proliferation of color indices, we resort 
to Greek superscripts)
\bea
\lefteqn{
\tilde{\G}_{\si;\ov{c}cc\ov{c}}^{adef}(p_1,p_2,p_3,p_4)=
gf^{abc}\int\dk{k}W_{\si\nu}^{b\nu}(k)W_{\ov{c}c}^{c\ga}(p_1-k)\times}
\nonumber\\&&
\left\{
\G_{\ov{c}c\nu}^{f\mu\nu}(p_4,-p_4-k,k)W_{\ov{c}c}^{\mu\e}(p_4+k)
\left[\G_{\ov{c}c\al}^{\e d\al}(p_4+k,p_2,-k-p_2-p_4)
W_{\al\ka}^{\al\ka}(k+p_2+p_4)\G_{\ov{c}c\ka}^{\ga e\ka}(p_1-k,p_3,k+p_2+p_4)
\right.\right.\nonumber\\&&\left.
-\G_{\ov{c}c\al}^{\e e\al}(p_4+k,p_3,p_1+p_2-k)
W_{\al\ka}^{\al\ka}(k-p_1-p_2)
\G_{\ov{c}c\ka}^{\ga d\ka}(p_1-k,p_2,k-p_1-p_2)\right]
\nonumber\\&&
+\G_{\ov{c}c\e}^{f\mu\e}(p_4,k+p_2,p_1+p_3-k)
W_{\ov{c}c}^{\mu\al}(-p_2-k)
\G_{\ov{c}c\nu}^{\al d\nu}(-p_2-k,p_2,k)W_{\e\ka}^{\e\ka}(k+p_2+p_4)
\G_{\ov{c}c\ka}^{\ga e\ka}(p_1-k,p_3,k+p_2+p_4)
\nonumber\\&&
-\G_{\ov{c}c\e}^{f\mu\e}(p_4,k+p_3,p_1+p_2-k)W_{\ov{c}c}^{\mu\al}(-p_3-k)
\G_{\ov{c}c\nu}^{\al e\nu}(-p_3-k,p_3,k)W_{\e\ka}^{\e\ka}(k+p_3+p_4)
\G_{\ov{c}c\ka}^{\ga d\ka}(p_1-k,p_2,k+p_3+p_4)
\nonumber\\&&
+\G_{\ov{c}c\al\nu}^{fe\al\nu}(p_4,p_3,-k-p_3-p_4,k)
W_{\al\ka}^{\al\ka}(k+p_3+p_4)\G_{\ov{c}c\ka}^{\ga d\ka}(p_1-k,p_2,k+p_3+p_4)
\nonumber\\&&
-\G_{\ov{c}c\al\nu}^{fd\al\nu}(p_4,p_2,-k-p_2-p_4,k)
W_{\al\ka}^{\al\ka}(k+p_2+p_4)\G_{\ov{c}c\ka}^{\ga e\ka}(p_1-k,p_3,k+p_2+p_4)
\nonumber\\&&
+\G_{\ov{c}c\ov{c}c}^{f\ka\ga d}(p_4,k+p_3,p_1-k,p_2)
W_{\ov{c}c}^{\ka\al}(-p_3-k)\G_{\ov{c}c\nu}^{\al e\nu}(-p_3-k,p_3,k)
\nonumber\\&&
-\G_{\ov{c}c\ov{c}c}^{f\ka\ga e}(p_4,k+p_2,p_1-k,p_3)
W_{\ov{c}c}^{\ka\al}(-p_2-k)\G_{\ov{c}c\nu}^{\al d\nu}(-p_2-k,p_2,k)
\nonumber\\&&\left.
+\G_{\ov{c}c\nu}^{f\mu\nu}(p_4,-p_4-k,k)W_{\ov{c}c}^{\mu\ka}(p_4+k)
\G_{\ov{c}c\ov{c}c}^{\ka e\ga d}(p_4+k,p_3,p_1-k,p_2)
-\G_{\ov{c}c\ov{c}c\nu}^{fe\ga d\nu}(p_4,p_3,p_1-k,p_2,k)
\right\},
\\
\lefteqn{
\tilde{\G}_{\ov{c};\ov{c}cc\si}^{adeg}(p_1,p_2,p_3,p_4)=
gf^{abc}\int\dk{k}W_{\ov{c}c}^{b\nu}(k)W_{\ov{c}c}^{c\ga}(p_1-k)
\times}
\nonumber\\&&
\left\{
2\G_{\ov{c}c\si}^{\nu\mu g}(k,-p_4-k,p_4)W_{\ov{c}c}^{\mu\e}(p_4+k)
\G_{\ov{c}c\al}^{\e e\al}(p_4+k,p_3,p_1+p_2-k)
W_{\al\ka}^{\al\ka}(k-p_1-p_2)\G_{\ov{c}c\ka}^{\ga d\ka}(p_1-k,p_2,k-p_1-p_2)
\right.\nonumber\\&&
-2\G_{\ov{c}c\si}^{\nu\mu g}(k,-p_4-k,p_4)
W_{\ov{c}c}^{\mu\e}(p_4+k)
\G_{\ov{c}c\al}^{\e d\al}(p_4+k,p_2,p_1+p_3-k)
W_{\al\ka}^{\al\ka}(k-p_1-p_3)\G_{\ov{c}c\ka}^{\ga e\ka}(p_1-k,p_3,k-p_1-p_3)
\nonumber\\&&
-2\G_{\ov{c}c\e\si}^{\nu e\e g}(k,p_3,-p_3-p_4-k,p_4)
\G_{\ov{c}c\ka}^{\ga d\ka}(p_1-k,p_2,k+p_3+p_4)W_{\e\ka}^{\e\ka}(k+p_3+p_4)
\nonumber\\&&
+2\G_{\ov{c}c\e\si}^{\nu d\e g}(k,p_2,-p_2-p_4-k,p_4)
\G_{\ov{c}c\ka}^{\ga e\ka}(p_1-k,p_3,k+p_2+p_4)W_{\e\ka}^{\e\ka}(k+p_2+p_4)
\nonumber\\&&
+2\G_{\ov{c}c\e}^{\nu e\e}(k,p_3,-p_3-k)W_{\e\mu}^{\e\mu}(p_3+k)
\G_{\mu\si\al}^{\mu g\al}(p_3+k,p_4,p_1+p_2-k)
W_{\al\ka}^{\al\ka}(k-p_1-p_2)\G_{\ov{c}c\ka}^{\ga d\ka}(p_1-k,p_2,k-p_1-p_2)
\nonumber\\&&\left.
-2\G_{\ov{c}c\si}^{\nu\e g}(k,-p_4-k,p_4)W_{\ov{c}c}^{\e\ka}(p_4+k)
\G_{\ov{c}c\ov{c}c}^{\ka e\ga d}(p_4+k,p_3,p_1-k,p_2)
+\G_{\ov{c}c\ov{c}c\si}^{\nu e\ga dg}(k,p_3,p_1-k,p_2,p_4)
\right\}.
\eea
(A similar expression exists for the kernel 
$\tilde{\G}_{A;\ov{c}cc\ov{c}}$.)  As before, since we may choose $q_0$ at 
will and the energy ($k_4^0$) is scalar, we now have two more expressions 
that are able to give $\G_{\ov{c}c\si A}$ and $\G_{\ov{c}c\si\si}$ in terms 
of all the other previously considered Green's functions involving the 
$\si$-field except one: $\G_{\ov{c}c\ov{c}c\si}$.  Also as before, one can 
immediately verify this identity at tree-level using the Feynman rules of 
Ref.~\cite{Watson:2007vc} (and repeated in Appendix~\ref{app:decomp}).

\subsection{Two pairs of further ghost derivatives}
In order to close the system of \ST identities, we must find an equation for 
$\G_{\ov{c}c\ov{c}c\si}$.  This equation arises by functionally 
differentiating \eq{eq:wtid3} with respect to $\imath c_w^e$, 
$\imath\ov{c}_v^f$, $\imath c_u^g$ and $\imath\ov{c}_r^h$ and then setting 
sources to zero.  Using the results Eqs.~(\ref{eq:tilde0}) and 
(\ref{eq:tilde1}) and the appropriate functional derivatives of 
\eq{eq:gheom0} we obtain \eq{eq:inter2} (the new kernel will be discussed 
below).  This expression is cyclic symmetric in the three ghost derivatives 
$\imath c_z^d$, $\imath c_w^e$ and $\imath c_u^g$ and is antisymmetric in 
the two derivatives $\imath\ov{c}_v^f$ and $\imath\ov{c}_r^h$.  In momentum 
space, the identity reads:
\bea
0&=&k_5^0\G_{\ov{c}c\ov{c}c\si}^{hgfde}(k_1,k_2,k_3,k_4,k_5)
-\frac{\imath k_{5i}}{\vec{k}_5^2}
\G_{\ov{c}c\ov{c}cAi}^{hgfda}(k_1,k_2,k_3,k_4,k_5)
\G_{\ov{c}c}^{ae}(-\vec{k}_5)
\nonumber\\&&
+(k_2^0+q_0)\G_{\ov{c}c\ov{c}c\si}^{hdfeg}(k_1,k_4,k_3,k_5-q_0,k_2+q_0)
-\frac{\imath k_{2i}}{\vec{k}_2^2}
\G_{\ov{c}c\ov{c}cAi}^{hdfea}(k_1,k_4,k_3,k_5-q_0,k_2+q_0)
\G_{\ov{c}c}^{ag}(-\vec{k}_2)
\nonumber\\&&
+(k_4^0+q_0)\G_{\ov{c}c\ov{c}c\si}^{hefgd}(k_1,k_5-q_0,k_3,k_2,k_4+q_0)
-\frac{\imath k_{4i}}{\vec{k}_4^2}
\G_{\ov{c}c\ov{c}cAi}^{hefga}(k_1,k_5-q_0,k_3,k_2,k_4+q_0)
\G_{\ov{c}c}^{ad}(-\vec{k}_4)
\nonumber\\&&
+\G_{\ov{c}c\si}^{fda}(k_3,k_4,-k_3-k_4)
\tilde{\G}_{\si;\ov{c}cc\ov{c}}^{aegh}(q_0+k_3+k_4,k_5-q_0,k_2,k_1)
\nonumber\\&&
+\G_{\ov{c}c\si}^{fea}(k_3,k_5-q_0,q_0-k_3-k_5)
\tilde{\G}_{\si;\ov{c}cc\ov{c}}^{agdh}(k_3+k_5,k_2,k_4,k_1)
\nonumber\\&&
+\G_{\ov{c}c\si}^{fga}(k_3,k_2,-k_2-k_3)
\tilde{\G}_{\si;\ov{c}cc\ov{c}}^{adeh}(q_0+k_2+k_3,k_4,k_5-q_0,k_1)
\nonumber\\&&
+\G_{\ov{c}cAi}^{fda}(k_3,k_4,-k_3-k_4)
\tilde{\G}_{A;\ov{c}cc\ov{c}i}^{aegh}(q_0+k_3+k_4,k_5-q_0,k_2,k_1)
\nonumber\\&&
+\G_{\ov{c}cAi}^{fea}(k_3,k_5-q_0,q_0-k_3-k_5)
\tilde{\G}_{A;\ov{c}cc\ov{c}i}^{agdh}(k_3+k_5,k_2,k_4,k_1)
\nonumber\\&&
+\G_{\ov{c}cAi}^{fga}(k_3,k_2,-k_2-k_3)
\tilde{\G}_{A;\ov{c}cc\ov{c}i}^{adeh}(q_0+k_2+k_3,k_4,k_5-q_0,k_1)
\nonumber\\&&
-\G_{\ov{c}c\si}^{hda}(k_1,k_4,-k_1-k_4)
\tilde{\G}_{\si;\ov{c}cc\ov{c}}^{aegf}(q_0+k_1+k_4,k_5-q_0,k_2,k_3)
\nonumber\\&&
-\G_{\ov{c}c\si}^{hea}(k_1,k_5-q_0,q_0-k_1-k_5)
\tilde{\G}_{\si;\ov{c}cc\ov{c}}^{agdf}(k_1+k_5,k_2,k_4,k_3)
\nonumber\\&&
-\G_{\ov{c}c\si}^{hga}(k_1,k_2,-k_1-k_2)
\tilde{\G}_{\si;\ov{c}cc\ov{c}}^{adef}(q_0+k_1+k_2,k_4,k_5-q_0,k_3)
\nonumber\\&&
-\G_{\ov{c}cAi}^{hda}(k_1,k_4,-k_1-k_4)
\tilde{\G}_{A;\ov{c}cc\ov{c}i}^{aegf}(q_0+k_1+k_4,k_5-q_0,k_2,k_3)
\nonumber\\&&
-\G_{\ov{c}cAi}^{hea}(k_1,k_5-q_0,q_0-k_1-k_5)
\tilde{\G}_{A;\ov{c}cc\ov{c}i}^{agdf}(k_1+k_5,k_2,k_4,k_3)
\nonumber\\&&
-\G_{\ov{c}cAi}^{hga}(k_1,k_2,-k_1-k_2)
\tilde{\G}_{A;\ov{c}cc\ov{c}i}^{adef}(q_0+k_1+k_2,k_4,k_5-q_0,k_3)
\nonumber\\&&
+\frac{1}{2}\G_{\ov{c}c\ov{c}c}^{hdfa}(k_1,k_4,k_3,k_2+k_5)
\left[\tilde{\G}_{\ov{c};\ov{c}cc}^{aeg}(q_0-k_2-k_5,k_5-q_0,k_2)
-2gf^{aeg}\right]
\nonumber\\&&
+\frac{1}{2}\G_{\ov{c}c\ov{c}c}^{hefa}(k_1,k_5-q_0,k_3,q_0+k_2+k_4)
\left[\tilde{\G}_{\ov{c};\ov{c}cc}^{agd}(-k_2-k_4,k_2,k_4)
-2gf^{agd}\right]
\nonumber\\&&
+\frac{1}{2}\G_{\ov{c}c\ov{c}c}^{hgfa}(k_1,k_2,k_3,k_4+k_5)
\left[\tilde{\G}_{\ov{c};\ov{c}cc}^{ade}(q_0-k_4-k_5,k_4,k_5-q_0)
-2gf^{ade}\right]
\nonumber\\&&
+\frac{1}{2}\G_{\ov{c}c}^{fa}(\vec{k}_3)
\tilde{\G}_{\ov{c};\ov{c}ccc\ov{c}}^{adegh}(q_0+k_3,k_4,k_5-q_0,k_2,k_1)
-\frac{1}{2}\G_{\ov{c}c}^{ha}(\vec{k}_1)
\tilde{\G}_{\ov{c};\ov{c}ccc\ov{c}}^{adegf}(q_0+k_1,k_4,k_5-q_0,k_2,k_3).
\label{eq:stid5}
\eea
Aside from the Green's function that we wish to calculate 
($\G_{\ov{c}c\ov{c}c\si}$), there is only one further unknown kernel: 
$\tilde{\G}_{\ov{c};\ov{c}ccc\ov{c}}$, given in \eq{eq:rnd4}.  Importantly, 
this kernel introduces no new Green's functions involving functional 
derivatives with respect to the $\si$-field and in momentum space, it reads:
\bea
\lefteqn{
\tilde{\G}_{\ov{c};\ov{c}ccc\ov{c}}^{adegf}(p_1,p_2,p_3,p_4,p_5)=
gf^{abc}\int\dk{k}W_{\ov{c}c}^{b\mu}(k)W_{\ov{c}c}^{c\nu}(p_1-k)
\left\{
-\frac{1}{3}\G_{\ov{c}c\ov{c}c\ov{c}c}^{\mu gfe\nu d}(k,p_4,p_5,p_3,p_1-k,p_2)
\right.}\nonumber\\&&
+2\G_{\ov{c}c\ov{c}c}^{f\ka\nu d}(p_5,k+p_3+p_4,p_1-k,p_2)
W_{\ov{c}c}^{\ka\e}(-k-p_3-p_4)
\G_{\ov{c}c\ov{c}c}^{\mu g\e e}(k,p_4,-k-p_3-p_4,p_3)
\nonumber\\&&
+2\G_{\ov{c}c\ov{c}c\ba}^{fd\nu g\ba}(p_5,p_2,p_1-k,p_4,k+p_3)
W_{\al\ba}^{\al\ba}(k+p_3)\G_{\ov{c}c\al}^{\mu e\al}(k,p_3,-k-p_3)
\nonumber\\&&
+2\G_{\ov{c}c\ga\ba}^{fd\ga\ba}(p_5,p_2,p_1+p_3-k,k+p_4)
\G_{\ov{c}c\al}^{\mu g\al}(k,p_4,-k-p_4)W_{\al\ba}^{\al\ba}(k+p_4)
W_{\ga\de}^{\ga\de}(k-p_1-p_3)\G_{\ov{c}c\de}^{\nu e\de}(p_1-k,p_3,k-p_1-p_3)
\nonumber\\&&
+2\G_{\ov{c}c\ba}^{f\ka\ba}(p_5,-k-p_2-p_5,k+p_2)
\G_{\ov{c}c\al}^{\mu d\al}(k,p_2,-k-p_2)
\G_{\ov{c}c\de}^{\nu e\de}(p_1-k,p_3,k-p_1-p_3)
\G_{\ov{c}c\ga}^{\e g\ga}(k+p_2+p_5,p_4,p_1+p_3-k)
\nonumber\\&&
\times W_{\al\ba}^{\al\ba}(k+p_2)W_{\ga\de}^{\ga\de}(k-p_1-p_3)
W_{\ov{c}c}^{\ka\e}(k+p_2+p_5)
\nonumber\\&&
+2\G_{\ov{c}c\ga}^{f\ka\ga}(p_5,k+p_2+p_4,p_1+p_3-k)
\G_{\ov{c}c\al}^{\mu d\al}(k,p_2,-k-p_2)
\G_{\ov{c}c\de}^{\nu e\de}(p_1-k,p_3,k-p_1-p_3)
\G_{\ov{c}c\ba}^{\e g\ba}(-k-p_2-p_4,p_4,k+p_2)
\nonumber\\&&
\times W_{\al\ba}^{\al\ba}(k+p_2)W_{\ga\de}^{\ga\de}(k-p_1-p_3)
W_{\ov{c}c}^{\ka\e}(-k-p_2-p_4)
\nonumber\\&&
-2\G_{\ov{c}c\ba}^{f\ka\ba}(p_5,-k-p_3-p_5,k+p_3)
\G_{\ov{c}c\ov{c}c}^{\e d\nu g}(k+p_3+p_5,p_2,p_1-k,p_4)
\G_{\ov{c}c\al}^{\mu e\al}(k,p_3,-k-p_3)
\nonumber\\&&
\times W_{\al\ba}^{\al\ba}(k+p_3)W_{\ov{c}c}^{\ka\e}(k+p_3+p_5)
\nonumber\\&&
+2\G_{\ov{c}c\ov{c}c}^{f\ka\nu d}(p_5,k+p_3+p_4,p_1-k,p_2)
\G_{\ov{c}c\al}^{\mu e\al}(k,p_3,-k-p_3)
\G_{\ov{c}c\ba}^{\e g\ba}(-k-p_3-p_4,p_4,k+p_3)
\nonumber\\&&
\times W_{\al\ba}^{\al\ba}(k+p_3)W_{\ov{c}c}^{\ka\e}(-k-p_3-p_4)
\nonumber\\&&
-2\G_{\ov{c}c\ov{c}c}^{f\ka\nu d}(p_5,k+p_3+p_4,p_1-k,p_2)
\G_{\ov{c}c\al}^{\mu g\al}(k,p_4,-k-p_4)
\G_{\ov{c}c\ba}^{\e e\ba}(-k-p_3-p_4,p_3,k+p_4)
\nonumber\\&&\left.
\times W_{\al\ba}^{\al\ba}(k+p_4)W_{\ov{c}c}^{\ka\e}(-k-p_3-p_4)
+\mbox{c.p. ($c^d(p_2),c^e(p_3),c^g(p_4)$)}\right\}
\eea
where we utilize the cyclic symmetry (cyclic symmetric terms denoted by 
$\mbox{c.p.}$ and this includes the first factor 
$\G_{\ov{c}c\ov{c}c\ov{c}c}$ which is the origin of the factor $1/3$).  
Because of the cyclic symmetry and the fact that $q_0$ is arbitrary, 
\eq{eq:stid5} can be solved for $\G_{\ov{c}c\ov{c}c\si}$ and we have finally 
managed to close the set of Slavnov--Taylor identities.  The identity, 
\eq{eq:stid5}, trivially has no tree-level form.

We have thus shown that just as for the temporal two-point functions, all 
temporal Green's functions considered so far can be deduced (at least in 
principle) from their non-temporal counterparts as solutions to the 
Slavnov--Taylor identities.  The temporal $\si$-field has been effectively 
eliminated from the system, or integrated out of functional form of the 
action.  Again, as for the two-point functions we see that the 
Slavnov--Taylor identities relate temporal, (spatially) longitudinal and 
ghost Green's functions in a manner reminiscent of the Kugo--Ojima quartet 
mechanism \cite{Kugo:1979gm}.

\subsection{Further Slavnov--Taylor identities}

It has been shown so far that there exists a closed set of Slavnov--Taylor 
identities that includes the vertex (three-point proper) Green's 
functions.  However, one may also consider four-point functions (e.g., 
$\G_{AAA\si}$).  Because of their extended nature, we do not attempt to 
derive these identities completely; rather, we shall merely sketch their 
form in order to justify that the equations close just as before.  We begin 
as previously with \eq{eq:wtid3} and again, the sets of derivatives are 
distinguished by how many pairs of ghost/antighost functional derivatives 
are taken.  We will highlight only those temporal Green's functions or 
kernels that have not previously appeared and leave those quantities that 
have already been derived as implicit.  The sequence is as follows:
\begin{enumerate}
\item
Taking functional derivatives of \eq{eq:wtid3} with respect to 
$\imath\Phi_\ka$, $\imath\Phi_\ta$ and $\imath\Phi_\la$ (as before, the 
ghost derivatives will be made explicit and the field types here refer only 
to either the $\vec{A}$ or $\si$-fields), one clearly obtains an identity 
for $\G_{\ka\ta\la\si}$ in terms of $\G_{\ka\ta\la A}$ and a new kernel 
$\tilde{\G}_{\si;\ov{c}c\la\ta}$.  This kernel, following (as a further 
functional derivative of) \eq{eq:rnd1} introduces the new quantity: 
$\G_{\ov{c}c\ka\ta\la}$.  Starting with $\G_{AAAA}$, we can then 
sequentially build up to $\G_{\si\si\si\si}$, given all the 
$\G_{\ov{c}c\ka\ta\la}$.
\item
Next we take one pair of ghost functional derivatives (i.e., functionally 
differentiate \eq{eq:wtid3} with respect to $\imath\ov{c}$ and $\imath c$) 
and then derivatives with respect to $\imath\Phi_\ta$ and 
$\imath\Phi_\la$.  This gives us equations for the $\G_{\ov{c}c\ta\la\si}$ 
in terms of the $\G_{\ov{c}c\ta\la A}$ and we have two new kernels: 
$\tilde{\G}_{\si;\ov{c}cc\ov{c}\la}$ and 
$\tilde{\G}_{\ov{c};\ov{c}cc\la\ta}$.  These kernels follow from 
Eqs.~(\ref{eq:rnd2}) and (\ref{eq:rnd3}), respectively, and both introduce 
the new function: $\G_{\ov{c}c\ov{c}c\ta\la}$.
\item
With two pairs of ghost functional derivatives of \eq{eq:wtid3}, plus one 
further derivative with respect to $\imath\Phi_\la$ we generate equations 
for $\G_{\ov{c}c\ov{c}c\la\si}$ in terms of $\G_{\ov{c}c\ov{c}c\la A}$, 
again with a new kernel: $\tilde{\G}_{\ov{c};\ov{c}ccc\ov{c}\la}$.  This 
kernel follows from \eq{eq:rnd4} and introduces a final Green's function: 
$\G_{\ov{c}c\ov{c}c\ov{c}c\la}$.
\item
We finally take three pairs of ghost functional derivatives to get an 
equation for $\G_{\ov{c}c\ov{c}c\ov{c}c\si}$ in terms of 
$\G_{\ov{c}c\ov{c}c\ov{c}cA}$.  However, this equation does not involve any 
further kernels and the set of equations terminates.
\end{enumerate}
Now, all these equations have the same characteristics as in the previous 
subsections: the energy is a scalar quantity and there is the energy 
injection scale $q_0$, such that one has a set of unambiguous (albeit 
nonlinear and extremely long) equations from which the temporal Green's 
functions may be derived, given the set of spatial and ghost Green's 
functions as external input.

We conjecture that in principle, the closed sets of Slavnov--Taylor 
identities in Coulomb gauge may be extended to include all higher 
$n$-point Green's functions and that they may be solved to give exact 
expressions for all the temporal Green's functions.  As may be appreciated 
though, how to prove such a general statement is not clear.  One may 
formulate some ideas based on the following observations:  Firstly, the 
scalar nature of the energy and the energy injection scale ($q_0$) will 
always be present in the Slavnov--Taylor identities of Coulomb gauge 
courtesy of the particular Gauss-BRST invariance of the theory (this 
certainly improves on the situation in linear covariant gauges where one 
has contractions of tensors such that transverse parts cannot be directly 
deduced).  Second, because ghost derivatives must come in pairs, each 
further pair reduces the number of possible non-ghost internal lines within 
the kernels -- the external ghost legs must be connected by an internal 
ghost propagator or a vertex with these two ghost legs.  This restricts the 
number of possible internal temporal Green's functions such that eventually 
the set of identities closes.  One can see the emergence of a characteristic 
pattern for the ghost functions necessary to form the kernels in the steps 
above: i.e., $(1.)~\G_{\ov{c}c\ka\ta\la}\rightarrow(2.)
~\G_{\ov{c}c\ov{c}c\ta\la}\rightarrow(3.)
~\G_{\ov{c}c\ov{c}c\ov{c}c\la}\rightarrow(4.)~0$.

\section{\label{sec:five}Simplifications of the identities}
\setcounter{equation}{0}

Given that the Slavnov--Taylor identities derived in previous sections 
relate the various types of Green's functions in an extended manner, it is 
pertinent to ask whether these relationships reduce in specific 
circumstances such that unambiguous statements can be made about the 
behavior of individual Green's functions, in particular the two-point 
functions.  The motivation is clear: such information (particularly in the 
infrared region) may be useful for determining how the confinement mechanism 
manifests itself (asymptotic freedom already being perturbatively 
established in the ultraviolet region of Coulomb gauge \cite{Watson:2007vc}) 
and also provides for input in other nonperturbative studies.  We argue here 
that unfortunately such information cannot be unambiguously extracted.

In the noncovariant Coulomb gauge, we must first sort out the temporal 
(energy) and spatial (momentum) scales.  The most obvious simplification of 
the Slavnov--Taylor identities concerns purely spacelike momenta where we 
know that if the postulate of Euclidicity is to hold (i.e., that the Wick 
rotation is valid), the Green's functions presumably can have no 
singularities.  One may in principle also consider timelike configurations, 
but this is certainly beyond the scope of the present analysis.  Thus, we 
must first set all energy scales to zero.  We define the infrared region as 
the limit as one or more of the momenta vanishes and we approach the 
light-cone, where singularities (or nontrivial zeroes) may appear.

Let us begin by considering the Slavnov--Taylor identities for the 
two-point functions: Eqs.~(\ref{eq:stid0}) and (\ref{eq:stid1}).  Using the 
general decompositions of Ref.~\cite{Watson:2007vc} (presented also in 
Appendix~\ref{app:decomp}) in terms of the (scalar) dressing functions, we 
get for general kinematical configurations:
\bea
-\imath k_0\vec{k}^2\G_{\si\si}(k_0^2,\vec{k}^2)&=&
-\imath k_0\vec{k}^2\G_{A\si}(k_0^2,\vec{k}^2)\G_{\ov{c}c}(\vec{k}^2),
\nonumber\\
\imath k_0^2k_k\G_{A\si}(k_0^2,\vec{k}^2)&=&
\imath k_0^2k_k\ov{\G}_{AA}(k_0^2,\vec{k}^2)\G_{\ov{c}c}(\vec{k}^2).
\eea
Since both equations have common kinematical prefactors, we can then write
\be
\G_{\si\si}(k_0^2,\vec{k}^2)
=\G_{A\si}(k_0^2,\vec{k}^2)\G_{\ov{c}c}(\vec{k}^2)
=\ov{\G}_{AA}(k_0^2,\vec{k}^2)\left[\G_{\ov{c}c}(\vec{k}^2)\right]^2.
\label{eq:2ptstid}
\ee
This holds for \emph{any} kinematical configuration and we conclude that 
the Slavnov--Taylor identities for the two-point functions alone can give 
no information about the value of the Green's functions, only the 
relationship between them.

Slightly less trivial is the Slavnov--Taylor identity for the ghost-gluon 
vertex, \eq{eq:stid3}.  Concentrating on spacelike momenta as discussed 
above, we see that neither of the $\G_{\ov{c}c\si}$ vertices will contribute 
when the energy is set to zero.  To study the equation, let us firstly 
decompose the ghost-gluon vertex into tree-level and dressed parts as 
follows:
\be
\G_{\ov{c}cAi}^{abc}(p_1,p_2,p_3)=-\imath gf^{abc}p_{1i}
-\imath gf^{abc} p_{1j}\tilde{\G}_{A;\ov{c}cAji}(p_1,p_2,p_3).
\label{eq:ghgldse}
\ee
Let us explain this decomposition.  Since we are dealing with a three-point 
function, the color factor ($f^{abc}$) and coupling ($g$) will always be 
common and can be extracted.  The appearance of $\tilde{\G}$ stems from the 
momentum space form of \eq{eq:ghgldse0} and in fact, given the definition 
\eq{eq:ghgldse1}, the above equation, \eq{eq:ghgldse}, is the \DS equation 
for the spatial ghost-gluon vertex.  As will be seen shortly, the 
contraction of $p_{1j}$ with $\tilde{\G}_{A;\ov{c}cAji}$ is important.  
Lastly, for the $\tilde{\G}_{\ov{c};\ov{c}cc}$ kernel, we can only extract 
the color and coupling factors and we write
\be
\tilde{\G}_{\ov{c};\ov{c}cc}^{abc}=gf^{abc}\tilde{\G}_{\ov{c};\ov{c}cc}
\ee
At zero energy, \eq{eq:stid3} in terms of dressing functions is thus
\bea
0&=&-\s{\vec{k}_1}{\vec{k}_3}\G_{\ov{c}c}(\vec{k}_3^2)
-\s{\vec{k}_1}{\vec{k}_2}\G_{\ov{c}c}(\vec{k}_2^2)
-\vec{k}_1^2\G_{\ov{c}c}(\vec{k}_1^2)
-k_{1j}k_{3i}\tilde{\G}_{A;\ov{c}cAji}(\vec{k}_1,\vec{k}_2,\vec{k}_3)
\G_{\ov{c}c}(\vec{k}_3^2)
-k_{1j}k_{2i}\tilde{\G}_{A;\ov{c}cAji}(\vec{k}_1,\vec{k}_3,\vec{k}_2)
\G_{\ov{c}c}(\vec{k}_2^2)\nonumber\\
&&+\frac{1}{2}\vec{k}_1^2\G_{\ov{c}c}(\vec{k}_1^2)
\tilde{\G}_{\ov{c};\ov{c}cc}(\vec{k}_1,\vec{k}_2,\vec{k}_3).
\label{eq:lim0}
\eea
This equation is directly analogous to the Landau gauge Slavnov--Taylor 
identity for the ghost-gluon vertex which was studied under the truncation 
$\tilde{\G}_{\ov{c};\ov{c}cc}=0$ \cite{von Smekal:1997vx}.  Quite generally, 
it states that the kernels ($\tilde{\G}_{A;\ov{c}cAji}$ and 
$\tilde{\G}_{\ov{c};\ov{c}cc}$) and the two-point ghost dressing function 
$\G_{\ov{c}c}$ are nontrivially related.

Let us now discuss the above identity, \eq{eq:lim0}, in the infrared to 
assess whether it simplifies further.  Because of the symmetry, there are 
two limits of interest: $k_1\rightarrow0$ and $k_2\rightarrow0$.  For the 
infrared limit $k_1\rightarrow0$, we see that the entire equation has an 
overall factor $\sim|k_1|$ (in fact, one can show that there is the overall 
factor $|k_1|^2$).  Therefore, in the limit, no single Green's function is 
isolated from which to determine a value for this kinematical 
configuration.  Further, the Green's functions or their combination may even 
be singular.  For the limit $k_2\rightarrow0$, we recall that in Coulomb 
gauge, ghost vertex dressing functions vanish as the ``in-ghost" momentum 
vanishes \cite{Watson:2006yq}, i.e.,
\bea
\tilde{\G}_{A;\ov{c}cAji}(k_1,k_2,k_3)&
\stackrel{\vec{k}_2\rightarrow0}{\propto}&|\vec{k}_2|,\nonumber\\
\tilde{\G}_{\ov{c};\ov{c}cc}(k_1,k_2,k_3)&
\stackrel{\vec{k}_2\rightarrow0}{\propto}&|\vec{k}_2|.
\label{eq:p2zero}
\eea
This is, of course, exactly the same as in Landau gauge 
\cite{Taylor:1971ff}.  Using the momentum conservation, $k_3=-k_1-k_2$, and 
we see that \eq{eq:lim0} has the overall factor $|k_2|$ as 
$k_2\rightarrow0$.  This leads to the conclusion that no information about 
an individual Green's function may be extracted in this limit either.  We 
conclude that on its own, the Slavnov--Taylor identity for the ghost-gluon 
vertex, \eq{eq:stid3}, does not yield unambiguous information about the 
Green's functions without further knowledge.

For the Slavnov--Taylor identities given by \eq{eq:stid2}, there is one 
special case where simplification occurs and this again results in a 
situation exactly analogous to Landau gauge.  Considering the $\G_{AA\si}$ 
identity (i.e., setting the external indices $\la$ and $\ta$ to be 
referring to $\vec{A}$-fields), one has explicitly
\bea
k_3^0\G_{AA\si lk}^{fed}(k_1,k_2,k_3)
&=&
\imath\frac{k_{3i}}{\vec{k}_3^2}\G_{3Alki}^{fea}(k_1,k_2,k_3)
\G_{\ov{c}c}^{ad}(-\vec{k}_3)
\nonumber\\&&
-\G_{AAki}^{ea}(k_2)
\left[\tilde{\G}_{A;\ov{c}cAil}^{adf}(k_2+q_0,k_3-q_0,k_1)
+gf^{adf}\de_{il}\right]
\nonumber\\&&
-\G_{AAli}^{fa}(k_1)
\left[\tilde{\G}_{A;\ov{c}cAik}^{ade}(k_1+q_0,k_3-q_0,k_2)
+gf^{ade}\de_{ik}\right]
\nonumber\\&&
-\G_{A\si k}^{ea}(k_2)\tilde{\G}_{\si;\ov{c}cAl}^{adf}(k_2+q_0,k_3-q_0,k_1)
\nonumber\\&&
-\G_{A\si l}^{fa}(k_1)\tilde{\G}_{\si;\ov{c}cAk}^{ade}(k_1+q_0,k_3-q_0,k_2).
\eea
Further setting all energy scales to zero and using the general 
decompositions ($\G_{3A}^{abc}=-\imath gf^{abc}\G_{3A}$, for the rest, see 
above or Appendix~\ref{app:decomp}) one has
\bea
k_{3i}\G_{3Alki}(\vec{k}_1,\vec{k}_2,\vec{k}_3)\G_{\ov{c}c}(\vec{k}_3^2)
&=&\vec{k}_2^2t_{kl}(\vec{k}_2)\G_{AA}(\vec{k}_2^2)
+\vec{k}_2^2t_{ki}(\vec{k}_2)\G_{AA}(\vec{k}_2^2)
\tilde{\G}_{A;\ov{c}cAil}(\vec{k}_2,\vec{k}_3,\vec{k}_1)
\nonumber\\
&&-\vec{k}_1^2t_{lk}(\vec{k}_1)\G_{AA}(\vec{k}_1^2)
-\vec{k}_1^2t_{li}(\vec{k}_1)\G_{AA}(\vec{k}_1^2)
\tilde{\G}_{A;\ov{c}cAik}(\vec{k}_1,\vec{k}_3,\vec{k}_2).
\eea
Taking the contraction of the above equation with $k_{2k}$, one has thus
\be
k_{2k}k_{3i}\G_{3Alki}(\vec{k}_1,\vec{k}_2,\vec{k}_3)=
-\vec{k}_1^2\frac{\G_{AA}(\vec{k}_1^2)}{\G_{\ov{c}c}(\vec{k}_3^2)}
t_{lj}(\vec{k}_1)
\left[k_{2j}
+k_{2k}\tilde{\G}_{A;\ov{c}cAjk}(\vec{k}_1,\vec{k}_3,\vec{k}_2)\right].
\ee
Now, the overall Bose--symmetry of the three-gluon vertex means that after 
extracting the color factor ($f^{abc}$), the dressing function $\G_{3A}$ is 
antisymmetric under exchange of any two legs and so, by interchanging 
$k_{2k}\leftrightarrow k_{3i}$ one can eliminate the three-gluon vertex to 
obtain an expression involving only the two-point ghost dressing function 
($\G_{\ov{c}c}$) and the kernel $\tilde{\G}_{A;\ov{c}cAjk}$ which, after 
canceling out the overall factors reads:
\be
0=t_{lj}(\vec{k}_1)\left\{k_{2j}\G_{\ov{c}c}(\vec{k}_2^2)
+k_{3j}\G_{\ov{c}c}(\vec{k}_3^2)+\G_{\ov{c}c}(\vec{k}_2^2)k_{2k}
\tilde{\G}_{A;\ov{c}cAjk}(\vec{k}_1,\vec{k}_3,\vec{k}_2)
+\G_{\ov{c}c}(\vec{k}_3^2)k_{3k}
\tilde{\G}_{A;\ov{c}cAjk}(\vec{k}_1,\vec{k}_2,\vec{k}_3)\right\}.
\label{eq:tripleA}
\ee
In Landau gauge, this identity is well-known \cite{Kim:1979ep,Ball:1980ax} 
and was considered in Ref.~\cite{Boucaud:2008ky} under reasonable 
assumptions to be indicative of an infrared finite ghost dressing function 
($\G_{\ov{c}c}$).  In the infrared limit $k_2\rightarrow0$ (equivalently 
for $k_3\rightarrow0$), we see that all terms have the prefactor $|k_2|$ 
since $k_{3j}t_{lj}(\vec{k}_1)=-k_{2j}t_{lj}(\vec{k}_1)$ and for the kernel 
we have the general infrared result \eq{eq:p2zero}.  Without further 
assumption, the following (conservative) statement is true of 
\eq{eq:tripleA}: if the \emph{particular contraction} above of the 
ghost-gluon vertex kernel ($\tilde{\G}_{A;\ov{c}cAjk}$) is vanishing, then 
the ghost dressing function is constant.  However, notice that the 
`particular contraction' above is not the same as that appearing in either 
the Slavnov--Taylor identity for the ghost-gluon vertex, \eq{eq:lim0} or 
the \DS equations from which $\G_{\ov{c}c}$ can be obtained.  Thus we 
conclude, as previously, that the above component, \eq{eq:tripleA}, of the 
Slavnov--Taylor identity, \eq{eq:stid2} does not yield unambiguous 
information about particular Green's functions without further input.

The special symmetric contraction case (above) of the three-gluon vertex 
Slavnov--Taylor identity aside, in order to say anything about the two-point 
functions from the identities \eq{eq:stid2}, one must know something about 
at least one of the vertices $\G_{\ta\la\si}$ or $\G_{\ta\la A}$.  Such 
information is not available from general considerations and so, one cannot 
make any simple statement.  Further, the rest of the Slavnov--Taylor 
identities, equations~(\ref{eq:stid4}) and (\ref{eq:stid5}), clearly involve 
too many higher $n$-point functions to have any hope of simplification.

From the above discussion, it seems clear that no information about the 
behavior of the two-point functions (or vertices) can be obtained from the 
Slavnov--Taylor identities alone.  As is obvious from the simple 
Slavnov--Taylor identities for the two-point functions summarized by 
\eq{eq:2ptstid}, one can only determine the relationship between the 
various Green's functions and this must be true for any kinematical 
configuration.  Our only unambiguous `outside' knowledge about the Green's 
functions was the peculiar infrared behavior of the ghost vertices, the 
absence of singularities for spacelike momenta and the Bose-symmetry of the 
three-gluon vertex (which is why the ghost-gluon vertex identity, 
\eq{eq:stid3}, and the identity \eq{eq:tripleA} were of particular 
interest), but even then, the dimensionality or complexity of the 
Slavnov--Taylor identities denied concrete conclusions.

\section{Summary, discussion and conclusions}
\setcounter{equation}{0}

In this paper, the Slavnov--Taylor identities for Coulomb gauge Yang--Mills 
theory within the second order formalism have been derived.  The starting 
point was the Gauss-BRST invariance of the action \cite{Zwanziger:1998ez}, 
characterized by the \emph{time-dependent} BRS variation and peculiar to 
Coulomb gauge.  It was found that for the two-point and vertex Green's 
functions, the Slavnov--Taylor identities form closed sets from which the 
temporal Green's functions can be unambiguously derived given the relevant 
spatial gluon and ghost Green's functions as input.  The extension of this 
to higher order Green's functions was also discussed.  Special cases of the 
identities were studied and it was seen that there is no simplification such 
that information about a particular Green's function could be extracted.

It is worthwhile at this stage to discuss the Coulomb gauge Slavnov--Taylor 
identities in their wider context since the connection between several 
different themes becomes apparent.  The first of these themes centers around 
Gauss' law.  In classical electrodynamics, Gauss' law allows one to 
determine the temporal component of the gauge field (i.e., $\si$) directly 
from the physical charge distribution.  In the canonical (Hamiltonian-based) 
formulation of Yang--Mills theory, Gauss' law is applied as an operator 
identity to define the physical state space and can be explicitly resolved 
in Coulomb gauge, yielding the so-called Coulomb term which comprises the 
confining properties of the theory \cite{Feuchter:2004mk}.  In the 
functional formalism, Gauss' law appears after the elimination of the 
temporal ($\si$) field from the action, either directly as in the case of 
the first order formalism \cite{Reinhardt:2008pr,Zwanziger:1998ez} or as 
here as the elimination from the effective action via the Slavnov--Taylor 
identities.  The resolution of Gauss' law and its connection to the physical 
state space is certainly less apparent in the functional formalism since 
Green's functions are not directly related to physical observables but this 
is compensated for by the observation that the cancellation of the 
gauge-dependent degrees of freedom (i.e., temporal, longitudinal and ghost) 
is manifest --- in the first order formalism one can reduce the functional 
integral to transverse spatial gluon degrees of freedom and the 
Slavnov--Taylor identities here express this explicitly for the Green's 
functions in local fashion.  However, the connection between the physical 
state space and the Green's functions is understood conceptually within 
the framework of the Kugo--Ojima confinement scenario \cite{Kugo:1979gm}: 
by postulating a well-defined BRS charge and physical state space, the 
cancellation of the gauge-dependent degrees of freedom followed and in 
Landau gauge led to the prediction that the ghost propagator is infrared 
enhanced.  In the Coulomb gauge functional formalism, we know explicitly 
that there exists at least a total charge that is conserved and vanishing 
\cite{Reinhardt:2008pr} which partly confirms the Kugo--Ojima postulate.  
This total charge arises from considering the temporal zero-modes inherent 
to Coulomb gauge and leads to the formal demonstration of the cancellation 
of the gauge dependent degrees of freedom.  In this study, we have shown 
that the Slavnov--Taylor identities stemming from the temporally nontrivial 
Gauss-BRST transform supply this cancellation in local fashion.  In this 
respect, the temporal features of the Coulomb gauge functional formalism can 
be seen to supply a link between the physical charge and states to the 
Green's functions of the underlying theory.

The temporally nontrivial nature of Coulomb gauge is manifested in two 
ways.  On the one hand, resolving the temporal zero-modes leads (as 
mentioned above) to the vanishing and conserved total charge 
\cite{Reinhardt:2008pr}; on the other hand, the extra temporal degree of 
freedom in the Gauss-BRST transform leads to the closure of the 
Slavnov--Taylor identities.  The temporal zero-modes of the Faddeev--Popov 
operator lead us to consider the connection to the Gribov--Zwanziger picture 
of confinement \cite{Gribov:1977wm,Zwanziger:1995cv,Zwanziger:1998ez}.  In 
this scenario, it is recognized that the zero-modes (which are induced by 
incomplete gauge-fixing) should be separated from the functional integration 
and via stochastic quantization and entropy arguments, the authors were 
able to show that the resulting spatial transverse gluon propagator would 
be suppressed in the infrared (and thus drops out from the physical 
spectrum) whereas the temporal propagator provided for a long-range 
confining force.  Just as with the Kugo--Ojima scenario, the ghost 
propagator in Landau gauge would also be infrared enhanced.  In the Coulomb 
gauge functional formalism insofar as the temporal zero-modes are 
concerned, one can see the parallels: their resolution leads to a total 
charge and the cancellation of gauge degrees of freedom, providing an 
explicit demonstration of selected features of both the Gribov--Zwanziger 
and Kugo--Ojima confinement scenarios.  However, one part is evidently 
missing --- a prediction for the infrared behavior of the ghost (or 
something similar).  Resolving the temporal zero-modes in Coulomb gauge 
restricted the functional integral to field configurations such that the 
total charge is conserved and vanishing, whereas the same temporal feature 
gave rise to Slavnov--Taylor identities that form closed sets but 
explicitly no information about a particular Green's function.  This 
underlies the quite general feature of functional techniques, namely, that 
in order to talk about Green's functions one must take functional 
derivatives and in this respect, the Slavnov--Taylor identities (and for 
that matter, the \DS equations too) represent functional differential 
equations.  Their solution is known only up to some `constant' of 
integration (for explicit examples of this, see 
Refs.~\cite{Epple:2007ut,Reinhardt:2008ij,Fischer:2008uz} and references 
therein).  In general this is not obvious since one necessarily has an 
infinite tower of equations to consider, but in the case of the Coulomb 
gauge Slavnov--Taylor identities derived here, the closure allows us to see 
clearly that this is precisely the case --- functional techniques lead to 
\emph{relations} between Green's functions.

Pragmatically, there are two natural avenues which to explore further.  
The first is to search for the `missing' prediction about the value of a 
specific Green's function (most likely the infrared behavior of the ghost) 
in order to complete the connection between the temporal aspects of Coulomb 
gauge and the two confinement scenarios: Kugo--Ojima and 
Gribov--Zwanziger.  The second direction to take is to use the 
Slavnov--Taylor identities here in order to construct a charge-conserving 
truncation scheme from which to solve the \DS equations of Coulomb gauge 
Yang--Mills theory, allowing for the eventuality that the solution may only 
be determined up to some external `boundary condition' in analogy to the 
solution of standard differential equations.  Both these directions are 
being pursued.

On a final note, one further consideration for the Slavnov--Taylor 
identities derived in this paper is their verification to one-loop order 
in perturbation theory (the tree-level forms are trivial).  This has in 
fact been done.  However, as can be appreciated from the lengthy 
expressions, such a technical exercise is not suitable for inclusion in the 
present paper.  The verification of the one-loop identities involves three 
components: the one-loop expansion of the vertex \DS equations (and the 
kernels presented in the text), the use of the inherent translational 
invariance of the loop integrals and identities for the color factors.  
Importantly, none of the loop integrals need be explicitly evaluated.

\begin{acknowledgments}
This work has been supported by the Deutsche Forschungsgemeinschaft (DFG) 
under contracts no. DFG-Re856/6-2,3.
\end{acknowledgments}
\appendix
\section{\label{app:decomp}Feynman rules and decompositions}
\setcounter{equation}{0}

For completeness, we present here selected results from 
Ref.~\cite{Watson:2007vc} concerning the tree-level forms and general 
decompositions of various Green's functions.  The general decompositions 
of the non-ghost propagators and proper two-point functions are given in 
Table~\ref{tab:decomp} (without the common color factor $\de^{ab}$).  For 
the ghosts, we have
\be
W_{\ov{c}c}^{ab}(k)=-\de^{ab}\frac{\imath}{\vec{k}^2}D_{\ov{c}c}(\vec{k}^2),
\;\;\;\;
\G_{\ov{c}c}^{ab}(k)=\de^{ab}\imath\vec{k}^2\G_{\ov{c}c}(\vec{k}^2).
\ee
At tree-level, all dressing functions are unity.  The tree-level vertices 
are given by (all momenta are defined as incoming)
\bea
\G_{\si AAjk}^{(0)abc}(p_a,p_b,p_c)&=&\imath gf^{abc}\de_{jk}(p_b^0-p_c^0),
\nonumber\\
\G_{\si A\si j}^{(0)abc}(p_a,p_b,p_c)&=&-\imath gf^{abc}(p_a-p_c)_j,
\nonumber\\
\G_{3A ijk}^{(0)abc}(p_a,p_b,p_c)&=&
-\imath gf^{abc}
\left[\de_{ij}(p_a-p_b)_k+\de_{jk}(p_b-p_c)_i+\de_{ki}(p_c-p_a)_j\right],
\nonumber\\
\G_{\ov{c}cA i}^{(0)abc}(p_{\ov{c}},p_c,p_A)&=&-\imath gf^{abc}p_{\ov{c}i}.
\eea

\begin{table}
\begin{tabular}{|c|c|c|}\hline
$W$&$A_j$&$\si$
\\\hline\rule[-2.4ex]{0ex}{5.5ex}
$A_i$&$t_{ij}(\vec{k})\frac{\imath}{(k_0^2-\vec{k}^2)}D_{AA}$&$0$
\\\hline\rule[-2.4ex]{0ex}{5.5ex}
$\si$&$0$&$\frac{\imath}{\vec{k}^2}D_{\si\si}$
\\\hline\rule[-2.4ex]{0ex}{5.5ex}
$\la$&$\frac{k_j}{\vec{k}^2}$&$\frac{k^0}{\vec{k}^2}D_{\si\la}$
\\\hline
\end{tabular}
\hspace{1cm}
\begin{tabular}{|c|c|c|}\hline
$\G$&$A_j$&$\si$
\\\hline\rule[-2.4ex]{0ex}{5.5ex}
$A_i$&$-\imath(k_0^2-\vec{k}^2)t_{ij}(\vec{k})\G_{AA}
-\imath k_0^2\frac{k_ik_j}{\vec{k}^2}\ov{\G}_{AA}$&
$\imath k^0k_i\G_{A\si}$
\\\hline\rule[-2.4ex]{0ex}{5.5ex}
$\si$&$\imath k^0k_j\G_{A\si}$&$-\imath\vec{k}^2\G_{\si\si}$
\\\hline\rule[-2.4ex]{0ex}{5.5ex}
$\la$&$-k_j$&$0$
\\\hline
\end{tabular}
\caption{\label{tab:decomp}General form of propagators [left] and two-point 
proper functions [right] (without color factors) in momentum space.  All 
dressing functions are functions of $k_0^2$ and $\vec{k}^2$.}
\end{table}

\section{\label{app:eqlist}List of lengthy formula}
In this appendix, we list lengthy (configuration space) formula, necessary 
for the derivation of the Slavnov--Taylor identities, that occur in the text.
\bea
0&=&\int\dx{x}\de(t-x_0)\left\{
\imath\left[\pd_x^0
\ev{\imath\ov{c}_v^f\imath c_z^d\imath\Phi_{\la(k)u}^g\imath\si_x^e}\right]
\de(w-x)
-\imath\left[\pd_x^0
\ev{\imath\ov{c}_v^f\imath c_w^e\imath\Phi_{\la(k)u}^g\imath\si_x^d}\right]
\de(z-x)
\right.\nonumber\\&&
-\left[\frac{\nabla_{ix}}{(-\nabla_x^2)}
\ev{\imath\ov{c}_v^f\imath c_z^d\imath\Phi_{\la(k)u}^g\imath A_{ix}^a}\right]
\ev{\imath\ov{c}_x^a\imath c_w^e}
+\left[\frac{\nabla_{ix}}{(-\nabla_x^2)}
\ev{\imath\ov{c}_v^f\imath c_w^e\imath\Phi_{\la(k)u}^g\imath A_{ix}^a}\right]
\ev{\imath\ov{c}_x^a\imath c_z^d}
\nonumber\\&&
+\ev{\imath\ov{c}_v^f\imath c_z^d\imath\si_x^a}
\left[\tilde{\G}_{\si;\ov{c}c\la(k)}^{aeg}(x,w,u)
-gf^{age}\de_{\si\la}\de(u-x)\de(w-x)\right]
\nonumber\\&&
-\ev{\imath\ov{c}_v^f\imath c_w^e\imath\si_x^a}
\left[\tilde{\G}_{\si;\ov{c}c\la(k)}^{adg}(x,z,u)
-gf^{agd}\de_{\si\la}\de(u-x)\de(z-x)\right]
\nonumber\\&&
+\ev{\imath\ov{c}_v^f\imath c_z^d\imath A_{ix}^a}
\left[\tilde{\G}_{A;\ov{c}c\la i(k)}^{aeg}(x,w,u)
-gf^{age}\de_{ki}\de_{A\la}\de(u-x)\de(w-x)\right]
\nonumber\\&&
-\ev{\imath\ov{c}_v^f\imath c_w^e\imath A_{ix}^a}
\left[\tilde{\G}_{A;\ov{c}c\la i(k)}^{adg}(x,z,u)
-gf^{agd}\de_{ki}\de_{A\la}\de(u-x)\de(z-x)\right]
\nonumber\\&&
+\frac{1}{2}\ev{\imath\ov{c}_v^f\imath c_x^a\imath\Phi_{\la(k)u}^g}
\left[\tilde{\G}_{\ov{c};\ov{c}cc}^{ade}(x,z,w)
-2gf^{ade}\de(z-x)\de(w-x)\right]
\nonumber\\&&\left.
-\ev{\imath\Phi_{\la(k)u}^g\imath\si_x^a}
\tilde{\G}_{\si;\ov{c}cc\ov{c}}^{adef}(x,z,w,v)
-\ev{\imath\Phi_{\la(k)u}^g\imath A_{ix}^a}
\tilde{\G}_{A;\ov{c}cc\ov{c}i}^{adef}(x,z,w,v)
+\frac{1}{2}\ev{\imath\ov{c}_v^f\imath c_x^a}
\tilde{\G}_{\ov{c};\ov{c}cc\la(k)}^{adeg}(x,z,w,u)
\right\}.
\nonumber\\&&
\label{eq:inter1}
\eea
\bea
\tilde{\G}_{\si;\ov{c}cc\ov{c}}^{adef}(x,z,w,v)
&=&
\left.gf^{abc}
\frac{\de^3}{\de\imath\ov{c}_v^f\de\imath c_w^e\de\imath c_z^d}
\ev{\imath\ro_x^b\imath\ov{\et}_x^c}\right|_{J=0}
\nonumber\\&=&
gf^{abc}\ev{\imath\ro_x^b\imath J_\nu}
\ev{\imath\ov{\et}_x^c\imath\et_\ga}\times\nonumber\\&&
\left\{
\ev{\imath\ov{c}_v^f\imath c_\mu\imath\Phi_\nu}
\ev{\imath\ov{\et}_\mu\imath\et_\e}\ev{\imath J_\al\imath J_\ka}
\left[\ev{\imath\ov{c}_\e\imath c_z^d\imath\Phi_\al}
\ev{\imath\ov{c}_\ga\imath c_w^e\imath\Phi_\ka}
-\ev{\imath\ov{c}_\e\imath c_w^e\imath\Phi_\al}
\ev{\imath\ov{c}_\ga\imath c_z^d\imath\Phi_\ka}\right]
\right.\nonumber\\&&
+\ev{\imath\ov{c}_v^f\imath c_\mu\imath\Phi_\e}
\ev{\imath\ov{\et}_\mu\imath\et_\al}\ev{\imath J_\e\imath J_\ka}
\left[\ev{\imath\ov{c}_\al\imath c_z^d\imath\Phi_\nu}
\ev{\imath\ov{c}_\ga\imath c_w^e\imath\Phi_\ka}
-\ev{\imath\ov{c}_\al\imath c_w^e\imath\Phi_\nu}
\ev{\imath\ov{c}_\ga\imath c_z^d\imath\Phi_\ka}\right]
\nonumber\\&&
+\ev{\imath\ov{c}_v^f\imath c_w^e\imath\Phi_\al\imath\Phi_\nu}
\ev{\imath J_\al\imath J_\ka}\ev{\imath\ov{c}_\ga\imath c_z^d\imath\Phi_\ka}
-\ev{\imath\ov{c}_v^f\imath c_z^d\imath\Phi_\al\imath\Phi_\nu}
\ev{\imath J_\al\imath J_\ka}\ev{\imath\ov{c}_\ga\imath c_w^e\imath\Phi_\ka}
\nonumber\\&&
+\ev{\imath\ov{c}_v^f\imath c_\ka\imath\ov{c}_\ga\imath c_z^d}
\ev{\imath\ov{\et}_\ka\imath\et_\al}
\ev{\imath\ov{c}_\al\imath c_w^e\imath\Phi_\nu}
-\ev{\imath\ov{c}_v^f\imath c_\ka\imath\ov{c}_\ga\imath c_w^e}
\ev{\imath\ov{\et}_\ka\imath\et_\al}
\ev{\imath\ov{c}_\al\imath c_z^d\imath\Phi_\nu}
\nonumber\\&&\left.
+\ev{\imath\ov{c}_v^f\imath c_\mu\imath\Phi_\nu}
\ev{\imath\ov{\et}_\mu\imath\et_\ka}
\ev{\imath\ov{c}_\ka\imath c_w^e\imath\ov{c}_\ga\imath c_z^d}
-\ev{\imath\ov{c}_v^f\imath c_w^e\imath\ov{c}_\ga\imath c_z^d\imath\Phi_\nu}
\right\}.
\label{eq:rnd2}\\
\tilde{\G}_{\ov{c};\ov{c}cc\la}^{adeg}(x,z,w,u)
&=&
\left.gf^{abc}
\frac{\de^3}{\de\imath\Phi_{\la u}^g\de\imath c_w^e\de\imath c_z^d}
\ev{\imath\ov{\et}_x^b\imath\ov{\et}_x^c}\right|_{J=0}
\nonumber\\&=&
gf^{abc}\ev{\imath\ov{\et}_x^b\imath\et_\nu}
\ev{\imath\ov{\et}_x^c\imath\et_\ga}\times\nonumber\\&&
\left\{
2\ev{\imath\ov{c}_\nu\imath c_\mu\imath\Phi_{\la u}^g}
\ev{\imath\ov{\et}_\mu\imath\et_\e}\ev{\imath J_\al\imath J_\ka}\!
\left[\ev{\imath\ov{c}_\e\imath c_w^e\imath\Phi_\al}
\ev{\imath\ov{c}_\ga\imath c_z^d\imath\Phi_\ka}
-\ev{\imath\ov{c}_\e\imath c_z^d\imath\Phi_\al}
\ev{\imath\ov{c}_\ga\imath c_w^e\imath\Phi_\ka}\right]
\right.\nonumber\\&&
-2\ev{\imath J_\e\imath\Phi_\ka}
\left[\ev{\imath\ov{c}_\nu\imath c_w^e\imath\Phi_\e\imath\Phi_{\la u}^g}
\ev{\imath\ov{c}_\ga\imath c_z^d\imath\Phi_\ka}
-\ev{\imath\ov{c}_\nu\imath c_z^d\imath\Phi_\e\imath\Phi_{\la u}^g}
\ev{\imath\ov{c}_\ga\imath c_w^e\imath\Phi_\ka}\right]
\nonumber\\&&
+2\ev{\imath\ov{c}_\nu\imath c_w^e\imath\Phi_\e}\ev{\imath J_\e\imath J_\mu}
\ev{\imath\Phi_\mu\imath\Phi_{\la u}^g\imath\Phi_\al}
\ev{\imath J_\al\imath J_\ka}\ev{\imath\ov{c}_\ga\imath c_z^d\imath\Phi_\ka}
\nonumber\\&&\left.
-2\ev{\imath\ov{c}_\nu\imath c_\e\imath\Phi_{\la u}^g}
\ev{\imath\ov{\et}_\e\imath\et_\ka}
\ev{\imath\ov{c}_\ka\imath c_w^e\imath\ov{c}_\ga\imath c_z^d}
+\ev{\imath\ov{c}_\nu\imath c_w^e\imath\ov{c}_\ga
\imath c_z^d\imath\Phi_{\la u}^g}
\right\}.
\label{eq:rnd3}
\eea
(In the above expression, \eq{eq:rnd3}, we omit the possible index ($k$) 
when the field type $\la$ refers to the $\vec{A}$-field for notational 
clarity.)
\bea
0&=&\int\dx{x}\de(t-x_0)\left\{
\left[\imath\pd_x^0
\ev{\imath\ov{c}_r^h\imath c_u^g\imath\ov{c}_v^f\imath c_z^d\imath\si_x^e}
\right]\de(w-x)
-\left[\frac{\nabla_{ix}}{(-\nabla_x^2)}
\ev{\imath\ov{c}_r^h\imath c_u^g\imath\ov{c}_v^f\imath c_z^d\imath A_{ix}^a}
\right]\ev{\imath\ov{c}_x^a\imath c_w^e}
\right.\nonumber\\&&
+\left[\imath\pd_x^0
\ev{\imath\ov{c}_r^h\imath c_z^d\imath\ov{c}_v^f\imath c_w^e\imath\si_x^g}
\right]\de(u-x)
-\left[\frac{\nabla_{ix}}{(-\nabla_x^2)}
\ev{\imath\ov{c}_r^h\imath c_z^d\imath\ov{c}_v^f\imath c_w^e\imath A_{ix}^a}
\right]\ev{\imath\ov{c}_x^a\imath c_u^g}
\nonumber\\&&
+\left[\imath\pd_x^0
\ev{\imath\ov{c}_r^h\imath c_w^e\imath\ov{c}_v^f\imath c_u^g\imath\si_x^d}
\right]\de(z-x)
-\left[\frac{\nabla_{ix}}{(-\nabla_x^2)}
\ev{\imath\ov{c}_r^h\imath c_w^e\imath\ov{c}_v^f\imath c_u^g\imath A_{ix}^a}
\right]\ev{\imath\ov{c}_x^a\imath c_z^d}
\nonumber\\&&
+\ev{\imath\ov{c}_v^f\imath c_z^d\imath\si_x^a}
\tilde{\G}_{\si;\ov{c}cc\ov{c}}^{aegh}(x,w,u,r)
-\ev{\imath\ov{c}_r^h\imath c_z^d\imath\si_x^a}
\tilde{\G}_{\si;\ov{c}cc\ov{c}}^{aegf}(x,w,u,v)
\nonumber\\&&
+\ev{\imath\ov{c}_v^f\imath c_w^e\imath\si_x^a}
\tilde{\G}_{\si;\ov{c}cc\ov{c}}^{agdh}(x,u,z,r)
-\ev{\imath\ov{c}_r^h\imath c_w^e\imath\si_x^a}
\tilde{\G}_{\si;\ov{c}cc\ov{c}}^{agdf}(x,u,z,v)
\nonumber\\&&
+\ev{\imath\ov{c}_v^f\imath c_u^g\imath\si_x^a}
\tilde{\G}_{\si;\ov{c}cc\ov{c}}^{adeh}(x,z,w,r)
-\ev{\imath\ov{c}_r^h\imath c_u^g\imath\si_x^a}
\tilde{\G}_{\si;\ov{c}cc\ov{c}}^{adef}(x,z,w,v)
\nonumber\\&&
+\ev{\imath\ov{c}_v^f\imath c_z^d\imath A_{ix}^a}
\tilde{\G}_{A;\ov{c}cc\ov{c}i}^{aegh}(x,w,u,r)
-\ev{\imath\ov{c}_r^h\imath c_z^d\imath A_{ix}^a}
\tilde{\G}_{A;\ov{c}cc\ov{c}i}^{aegf}(x,w,u,v)
\nonumber\\&&
+\ev{\imath\ov{c}_v^f\imath c_w^e\imath A_{ix}^a}
\tilde{\G}_{A;\ov{c}cc\ov{c}i}^{agdh}(x,u,z,r)
-\ev{\imath\ov{c}_r^h\imath c_w^e\imath A_{ix}^a}
\tilde{\G}_{A;\ov{c}cc\ov{c}i}^{agdf}(x,u,z,v)
\nonumber\\&&
+\ev{\imath\ov{c}_v^f\imath c_u^g\imath A_{ix}^a}
\tilde{\G}_{A;\ov{c}cc\ov{c}i}^{adeh}(x,z,w,r)
-\ev{\imath\ov{c}_r^h\imath c_u^g\imath A_{ix}^a}
\tilde{\G}_{A;\ov{c}cc\ov{c}i}^{adef}(x,z,w,v)\
\nonumber\\&&
+\frac{1}{2}\ev{\imath\ov{c}_r^h\imath c_z^d\imath\ov{c}_v^f\imath c_x^a}
\left[\tilde{\G}_{\ov{c};\ov{c}cc}^{aeg}(x,w,u)
-2gf^{aeg}\de(w-x)\de(u-x)\right]
\nonumber\\&&
+\frac{1}{2}\ev{\imath\ov{c}_r^h\imath c_w^e\imath\ov{c}_v^f\imath c_x^a}
\left[\tilde{\G}_{\ov{c};\ov{c}cc}^{agd}(x,u,z)
-2gf^{agd}\de(u-x)\de(z-x)\right]
\nonumber\\&&
+\frac{1}{2}\ev{\ov{c}_r^h\imath c_u^g\imath\ov{c}_v^f\imath c_x^a}
\left[\tilde{\G}_{\ov{c};\ov{c}cc}^{ade}(x,z,w)
-2gf^{ade}\de(z-x)\de(w-x)\right]
\nonumber\\&&\left.
+\frac{1}{2}\ev{\ov{c}_v^f\imath c_x^a}
\tilde{\G}_{\ov{c};\ov{c}ccc\ov{c}}^{adegh}(x,z,w,u,r)
-\frac{1}{2}\ev{\ov{c}_r^h\imath c_x^a}
\tilde{\G}_{\ov{c};\ov{c}ccc\ov{c}}^{adegf}(x,z,w,u,v)
\right\}.
\label{eq:inter2}
\eea
\bea
\lefteqn{\tilde{\G}_{\ov{c};\ov{c}ccc\ov{c}}^{adegf}(x,z,w,u,v)=
\left.gf^{abc}\frac{\de^4}{\de\imath\ov{c}_v^f\de\imath c_u^g
\de\imath c_w^e\de\imath c_z^d}
\ev{\imath\ov{\et}_x^b\ov{\et}_x^c}\right|_{J=0}}
\nonumber\\&=&
gf^{abc}\ev{\imath\ov{\et}_x^b\imath\et_\mu}
\ev{\imath\ov{\et}_x^c\imath\et_\nu}
\left\{
-\ev{\imath\ov{c}_\mu\imath c_u^g\ov{c}_v^f\imath c_w^e
\imath\ov{c}_\nu\imath c_z^d}
\right.\nonumber\\&&
+2\ev{\imath\ov{\et}_\ka\imath\et_\e}
\left[\ev{\imath\ov{c}_v^f\imath c_\ka\imath\ov{c}_\nu\imath c_z^d}
\ev{\imath\ov{c}_\mu\imath c_u^g\imath\ov{c}_\e\imath c_w^e}
+\mbox{c.p. ($c_z^d,c_w^e,c_u^g$)}\right]
\nonumber\\&&
+2\ev{\imath J_\al\imath J_\ba}
\left[
\ev{\imath\ov{c}_v^f\imath c_z^d\imath\ov{c}_\nu\imath c_u^g\imath\Phi_\ba}
\ev{\imath\ov{c}_\mu\imath c_w^e\imath\Phi_\al}
+\mbox{c.p. ($c_z^d,c_w^e,c_u^g$)}\right]
\nonumber\\&&
+2\ev{\imath J_\al\imath J_\ba}\ev{\imath J_\ga\imath J_\de}
\left[\ev{\imath\ov{c}_v^f\imath c_z^d\imath\Phi_\ga\imath\Phi_\ba}
\ev{\imath\ov{c}_\mu\imath c_u^g\imath\Phi_\al}
\ev{\imath\ov{c}_\nu\imath c_w^e\imath\Phi_\de}
+\mbox{c.p. ($c_z^d,c_w^e,c_u^g$)}\right]
\nonumber\\&&
+2\ev{\imath\ov{\et}_\ka\imath\et_\e}\ev{\imath J_\al\imath J_\ba}
\ev{\imath J_\ga\imath J_\de}\ev{\imath\ov{c}_v^f\imath c_\ka\imath\Phi_\ba}
\left[
\ev{\imath\ov{c}_\mu\imath c_z^d\imath\Phi_\al}
\ev{\imath\ov{c}_\nu\imath c_w^e\imath\Phi_\de}
\ev{\imath\ov{c}_\e\imath c_u^g\imath\Phi_\ga}
+\mbox{c.p. ($c_z^d,c_w^e,c_u^g$)}\right]
\nonumber\\&&
+2\ev{\imath\ov{\et}_\ka\imath\et_\e}\ev{\imath J_\al\imath J_\ba}
\ev{\imath J_\ga\imath J_\de}\ev{\imath\ov{c}_v^f\imath c_\ka\imath\Phi_\ga}
\left[
\ev{\imath\ov{c}_\mu\imath c_z^d\imath\Phi_\al}
\ev{\imath\ov{c}_\nu\imath c_w^e\imath\Phi_\de}
\ev{\imath\ov{c}_\e\imath c_u^g\imath\Phi_\ba}
+\mbox{c.p. ($c_z^d,c_w^e,c_u^g$)}\right]
\nonumber\\&&
-2\ev{\imath\ov{\et}_\ka\imath\et_\e}\ev{\imath J_\al\imath J_\ba}
\ev{\ov{c}_v^f\imath c_\ka\imath\Phi_\ba}
\left[\ev{\imath\ov{c}_\e\imath c_z^d\imath\ov{c}_\nu\imath c_u^g}
\ev{\imath\ov{c}_\mu\imath c_w^e\imath\Phi_\al}
+\mbox{c.p. ($c_z^d,c_w^e,c_u^g$)}\right]
\nonumber\\&&
+2\ev{\imath\ov{\et}_\ka\imath\et_\e}\ev{\imath J_\al\imath J_\ba}\times
\nonumber\\&&\left.
\left[
\ev{\imath\ov{c}_v^f\imath c_\ka\imath\ov{c}_\nu\imath c_z^d}
\left(
\ev{\imath\ov{c}_\mu\imath c_w^e\imath\Phi_\al}
\ev{\imath\ov{c}_\e\imath c_u^g\imath\Phi_\ba}
-\ev{\imath\ov{c}_\mu\imath c_u^g\imath\Phi_\al}
\ev{\imath\ov{c}_\e\imath c_w^e\imath\Phi_\ba}
\right)
+\mbox{c.p. ($c_z^d,c_w^e,c_u^g$)}\right]
\right\}.
\label{eq:rnd4}
\eea
Note that in this expression, terms cyclic symmetric in the three ghost 
derivatives $\imath c_z^d$, $\imath c_w^e$ and $\imath c_u^g$ are denoted 
\mbox{c.p. ($c_z^d,c_w^e,c_u^g$)}.



\begin{thebibliography}{99}

\bibitem{Slavnov:1972fg}
  A.~A.~Slavnov,
  Theor.\ Math.\ Phys.\  {\bf 10}, 99 (1972)
  [Teor.\ Mat.\ Fiz.\  {\bf 10}, 153 (1972)].

\bibitem{Taylor:1971ff}
  J.~C.~Taylor,
  Nucl.\ Phys.\  B {\bf 33}, 436 (1971).

\bibitem{Ball:1980ay}
  J.~S.~Ball and T.~W.~Chiu,
  Phys.\ Rev.\  D {\bf 22}, 2542 (1980).

\bibitem{Curtis:1990zs}
  D.~C.~Curtis and M.~R.~Pennington,
  Phys.\ Rev.\  D {\bf 42}, 4165 (1990).

\bibitem{Pennington:1998cj}
  M.~R.~Pennington,
  arXiv:hep-th/9806200.

\bibitem{Alkofer:2000wg}
  R.~Alkofer and L.~von Smekal,
  Phys.\ Rept.\  {\bf 353}, 281 (2001)
  [arXiv:hep-ph/0007355].

\bibitem{Kummer:1974ze}
  W.~Kummer,
  Acta Phys.\ Austriaca {\bf 41}, 315 (1975).

\bibitem{Baker:1980gf}
  M.~Baker, J.~S.~Ball and F.~Zachariasen,
  Nucl.\ Phys.\  B {\bf 186}, 531 (1981).

\bibitem{West:1982gg}
  G.~B.~West,
  Phys.\ Rev.\  D {\bf 27}, 1878 (1983).

\bibitem{Mandelstam:1979xd}
  S.~Mandelstam,
  Phys.\ Rev.\  D {\bf 20}, 3223 (1979).

\bibitem{BarGadda:1979cz}
  U.~Bar-Gadda,
  Nucl.\ Phys.\  B {\bf 163}, 312 (1980).

\bibitem{von Smekal:1997vx}
  L.~von Smekal, A.~Hauck and R.~Alkofer,
  Annals Phys.\  {\bf 267}, 1 (1998)
  [Erratum-ibid.\  {\bf 269}, 182 (1998)]
  [arXiv:hep-ph/9707327].

\bibitem{Fischer:2006ub}
  C.~S.~Fischer,
  J.\ Phys.\ G {\bf 32}, R253 (2006)
  [arXiv:hep-ph/0605173].

\bibitem{Watson:1999ha}
  P.~Watson,
  arXiv:hep-ph/9901454.

\bibitem{Watson:2001yv}
  P.~Watson and R.~Alkofer,
  Phys.\ Rev.\ Lett.\  {\bf 86}, 5239 (2001)
  [arXiv:hep-ph/0102332].

\bibitem{Schleifenbaum:2004id}
  W.~Schleifenbaum, A.~Maas, J.~Wambach and R.~Alkofer,
  Phys.\ Rev.\  D {\bf 72}, 014017 (2005)
  [arXiv:hep-ph/0411052].

\bibitem{Oehme:1979ai}
  R.~Oehme and W.~Zimmermann,
  Phys.\ Rev.\  D {\bf 21}, 471 (1980).

\bibitem{Kugo:1979gm}
  T.~Kugo and I.~Ojima,
  Prog.\ Theor.\ Phys.\ Suppl.\  {\bf 66}, 1 (1979);
  N.~Nakanishi and I.~Ojima,
  World Sci.\ Lect.\ Notes Phys.\  {\bf 27}, 1 (1990).

\bibitem{Gribov:1977wm}
  V.~N.~Gribov,
  Nucl.\ Phys.\ B {\bf 139} (1978) 1.

\bibitem{Zwanziger:1995cv}
  D.~Zwanziger,
  Nucl.\ Phys.\  B {\bf 485}, 185 (1997)
  [arXiv:hep-th/9603203].

\bibitem{Zwanziger:1998ez}
  D.~Zwanziger,
  Nucl.\ Phys.\ B {\bf 518} (1998) 237.

\bibitem{Boucaud:2008ky}
  Ph.~Boucaud, J.~P.~Leroy, A.~Le Yaouanc, J.~Micheli, O.~Pene and 
  J.~Rodriguez-Quintero,
  JHEP {\bf 0806}, 099 (2008)
  [arXiv:0803.2161 [hep-ph]].

\bibitem{Fischer:2008uz}
  C.~S.~Fischer, A.~Maas and J.~M.~Pawlowski,
  arXiv:0810.1987 [hep-ph].

\bibitem{Epple:2007ut}
  D.~Epple, H.~Reinhardt, W.~Schleifenbaum and A.~P.~Szczepaniak,
  Phys.\ Rev.\  D {\bf 77}, 085007 (2008)
  [arXiv:0712.3694 [hep-th]].

\bibitem{Feuchter:2004mk}
  C.~Feuchter and H.~Reinhardt,
  Phys.\ Rev.\ D {\bf 70} (2004) 105021
  [arXiv:hep-th/0408236];
  C.~Feuchter and H.~Reinhardt,
  arXiv:hep-th/0402106.

\bibitem{Abers:1973qs}
  E.~S.~Abers and B.~W.~Lee,
  Phys.\ Rept.\  {\bf 9}, 1 (1973).

\bibitem{Watson:2006yq}
  P.~Watson and H.~Reinhardt,
  Phys.\ Rev.\  D {\bf 75}, 045021 (2007)
  [arXiv:hep-th/0612114].

\bibitem{Watson:2007vc}
  P.~Watson and H.~Reinhardt,
  Phys.\ Rev.\  D {\bf 77}, 025030 (2008)
  [arXiv:0709.3963 [hep-th]].

\bibitem{Watson:2007mz}
  P.~Watson and H.~Reinhardt,
  Phys.\ Rev.\  D {\bf 76}, 125016 (2007)
  [arXiv:0709.0140 [hep-th]].

\bibitem{Popovici:2008ty}
  C.~Popovici, P.~Watson and H.~Reinhardt,
  arXiv:0810.4887 [hep-th].

\bibitem{Watson:2007fm}
  P.~Watson and H.~Reinhardt,
  arXiv:0711.2997 [hep-th].

\bibitem{Reinhardt:2008pr}
  H.~Reinhardt and P.~Watson,
  arXiv:0808.2436 [hep-th].

\bibitem{IZ}
  C.~Itzykson and J.~B.~Zuber, ``Quantum Field Theory", New York, USA: 
Mcgraw-Hill (1980) 705 P.(International Series in Pure and Applied Physics).

\bibitem{Marciano:1977su}
  W.~J.~Marciano and H.~Pagels,
  Phys.\ Rept.\  {\bf 36}, 137 (1978).

\bibitem{Kim:1979ep}
  S.~K.~Kim and M.~Baker,
  Nucl.\ Phys.\  B {\bf 164}, 152 (1980).

\bibitem{Ball:1980ax}
  J.~S.~Ball and T.~W.~Chiu,
  Phys.\ Rev.\  D {\bf 22}, 2550 (1980)
  [Erratum-ibid.\  D {\bf 23}, 3085 (1981)].

\bibitem{Reinhardt:2008ij}
  H.~Reinhardt and W.~Schleifenbaum,
  arXiv:0809.1764 [hep-th].

\end{thebibliography}
\end{document}